\documentclass[11pt,a4paper]{article} 
\usepackage[english,german]{babel}
\usepackage[latin1]{inputenc}
\usepackage[T1]{fontenc}
\usepackage{ae,aecompl}
\usepackage{times}
\usepackage{amsmath}
\usepackage{amsfonts,bbm,theorem}

\ifx\pdftexversion\undefined
  \usepackage[dvips]{graphicx}
\else
  \usepackage[pdftex]{graphicx}
\fi

\numberwithin{equation}{section}

\newtheorem{theorem}{Theorem}[section]
\newtheorem{definition}[theorem]{Definition}
\newtheorem{corollary}[theorem]{Corollary}
\newtheorem{assumption}[theorem]{Assumption}
\newtheorem{lemma}[theorem]{Lemma}

\newtheorem{proposition}[theorem]{Proposition}
\newenvironment{proof}{\begin{trivlist}\item[]{\em Proof:}\/}{%
\hfill\mbox{$\Box$}\end{trivlist}}
\newenvironment{proofof}[1]{\begin{trivlist}\item[]{\em Proof of #1:}\/}{%
\hfill\mbox{$\Box$}\end{trivlist}}

\newcommand{\barg}{\tilde{g}}
\newcommand{\torusd}{d_{\rm T}}
\newcommand{\tinj}[1]{[#1]'}
\newcommand{\Mf}{M_1}
\newcommand{\defem}[1]{{\em #1\/}}
\newcommand{\scint}{I_{\text{scr}}}
\newcommand{\sabs}[1]{\langle #1\rangle}

\newcommand{\vep}{\varepsilon}
\newcommand{\qand}{\qquad\text{and}\qquad}
\newcommand{\defset}[2]{ \left\{ #1\left|\, #2 %
      \makebox[0cm]{$\displaystyle\phantom{#1}$}\right.\!\right\} }
\newcommand{\set}[1]{\{#1\}}
\newcommand{\ci}{{\rm i}}
\newcommand{\rmd}{{\rm d}}

\newcommand{\1}{\mathbbm{1}}
\newcommand{\re}{{\rm Re\,}}

\newcommand{\Z}{{\mathbb Z}}
\newcommand{\R}{{\mathbb R}}
\newcommand{\T}{{\mathbb T}}
\newcommand{\C}{{\mathbb C\hspace{0.05 ex}}}
\newcommand{\N}{{\mathbb N}}
\newcommand{\frechet}{Fr\'{e}chet}
\newcommand{\norm}[1]{\Vert #1\Vert}

\begin{document}
\selectlanguage{english}

\newcommand{\email}[1]{E-mail: \tt #1}
\newcommand{\emailjani}{\email{jlukkari@ma.tum.de}}
\newcommand{\addressjani}{\em Zentrum Mathematik, 
Technische Universit\"at M\"unchen, \\
\em Boltzmannstr. 3, D-85747 Garching, Germany}

\title{Asymptotics of resolvent integrals:\\
The suppression of crossings for analytic lattice dispersion
relations }
\author{Jani Lukkarinen\thanks{\emailjani}\\[1em]
\addressjani }

\maketitle

\begin{abstract}
We study the so called crossing estimate for analytic
dispersion relations of periodic lattice systems 
in dimensions three and higher.  Under a certain regularity assumption on
the behavior of the dispersion relation near its critical values, we prove
that an analytic dispersion relation suppresses crossings if and only if it
is not a constant on any affine hyperplane.  In particular, this will 
then be true
for any dispersion relation which is a Morse function.
We also provide two examples of simple lattice systems whose dispersion
relations do not suppress crossings in the present sense.
\end{abstract}

\tableofcontents

\section{Introduction}
\label{sec:intro}

Time-dependent perturbation theory has proven 
to be a useful tool in studying the behavior of systems where a free,
wave-like evolution in three dimensions is
perturbed by a weak, random potential.  An important set of
tools for rigorous estimation of such a perturbation series
was developed by Erd\H{o}s and Yau  in \cite{erdyau99} to study the kinetic
limit of the random Schr\"{o}dinger evolution.  
These methods have later been extended to cover also
the low density limit of the random Schr\"{o}dinger evolution
\cite{eng04}, as well as the kinetic limits of 
an electron coupled to a phonon field \cite{erdos02},
of the discrete random Schr\"{o}dinger equation (the Anderson model)
\cite{chen03,chen04}, and of certain discrete wave equations with a weak,
random mass-disorder \cite{ls05}.
There is also a recent, remarkable result where the methods have been 
reworked to allow going beyond the kinetic time-scales for the continuum
and discrete random Schr\"{o}dinger evolution
\cite{erdyau04,erdyau05a,erdyau05b}.

An important element in all of these results is an estimate proving that
all so called {\em crossing graphs\/} are suppressed.  For the discrete
random Schr\"{o}dinger equation this was proven in \cite{chen03} 
by showing that for every sufficiently small $\beta>0$,
\begin{align}\label{eq:RSEsuppr}
&  \sup_{\alpha\in \R^3,k_0\in \T}\int_{(\T^3)^2} \rmd k_1\rmd k_2\, 
\frac{1}{|\alpha_1-\omega(k_1)+\ci\beta|
|\alpha_2-\omega(k_2)+\ci\beta|} 
\nonumber \\ & \quad\times
\frac{1}{|\alpha_3-\omega(k_1-k_2+k_0)+\ci\beta|} \le 
 c_1 \sabs{\ln \beta}^{n_1} \beta^{\gamma-1}.
\end{align}
Here $\omega(k) = \sum_{\nu=1}^3 (1- \cos(2\pi k_\nu))$ is 
{\em the dispersion relation \/} of the free, discrete random
Schr\"{o}dinger evolution, and
$c_1>0$, $n_1\ge 0$ and $\gamma>0$ are constants depending only on
the function $\omega$. 

We call (\ref{eq:RSEsuppr}) the {\em crossing estimate.\/}
The validity of the corresponding estimate in the earlier
continuum Schr\"{o}dinger case (when $\omega(k)=\frac{1}{2}k^2$, 
$k\in\R^3$)
was fairly easy to prove, but the proof
turned out to be involved in the discrete case,
$\omega(k) = \sum_{\nu=1}^3 (1- \cos(2\pi k_\nu))$.
There are now two independent proofs of this result: the bound in 
(\ref{eq:RSEsuppr})
was shown to hold with 
$\gamma=1/5$ and $n_1=2$ in Lemma 3.11 of \cite{chen03} 
and with $\gamma=1/4$ and $n_1=6$ in Appendix A.3 of \cite{erdyau04}.
The case of more general dispersion relations $\omega$
is not covered by the earlier results.  However,
in a very recent preprint by Erd\H{o}s and Salmhofer \cite{ES06},
the subject has been approached with a method
differing from ours.

For very small $\beta$, each of the factors in (\ref{eq:RSEsuppr}) 
are very sharply concentrated around some level sets of $\omega$.
However, the arguments of $\omega$ in the factors
are not allowed to vary independently of each other, and the magnitude of
the integral for small $\beta$ is thus determined by the overlap 
of the different level sets determined by the constants $\alpha_j$.
Therefore, to prove (\ref{eq:RSEsuppr}) it will be necessary 
to consider the worst case scenario for the level sets, and then try to
estimate the overlap between the three levels sets as $k_1$ and $k_2$ are
varied. 

However, it is not obvious how to carry out such an argument in the general
setup.  This raises the question:
for what kind of dispersion relations $\omega$ 
is it possible to derive the estimate (\ref{eq:RSEsuppr})?
This question is particularly relevant in the context of 
microscopic models for lattice vibrations in a crystal
where the dispersion relation is determined by the elastic couplings, and
can be fairly arbitrary (we refer to the survey \cite{spohn05},
and to its references, for further details on the topic).  
In an earlier work \cite{ls05}, where the perturbation
methods were applied to a simplified model of the lattice vibrations, 
the estimate (\ref{eq:RSEsuppr}) was in fact elevated to an assumption,
denoted by (DR4) in the paper.

Here our main aim is to show that the technical assumption (DR4) of the
earlier work \cite{ls05} can be replaced by a much simpler geometric
condition.  We will introduce the problem in detail and present the main
results in Section \ref{sec:main}, with the main notations collected to
Subsection \ref{sec:defs}.  Before proceeding to the more involved
proof of validity of the crossing estimate, we first prove the converse and
discuss a few counterexamples in Section \ref{sec:counterexamples}.
The proofs of the main theorems have been divided into Sections 
\ref{sec:techlemmas}--\ref{sec:crossing}.  Section \ref{sec:techlemmas}
collects the main technical lemmas, with some of the more well-known
details being reproduced for the sake of completeness in 
Appendices \ref{sec:composite} and \ref{sec:sabsxprop}. 
We prove in Section 
\ref{sec:proofofsemi} that the technical assumption made about the nature
of the set of singular points 
of the dispersion relation leads to a property similar to the usual
dispersivity. To show that the assumptions are fairly general, we 
have also included  in Appendix \ref{sec:Morse} a proof which shows 
that real-analytic Morse functions are covered by the main theorems.
The proof of the suppression of crossings is 
the content of Section \ref{sec:crossing}, where the first part gives a certain
uniform estimate on the minimal curvature of $\omega$ and the second part
exploits this to provide for the extra decay of the crossing integral.

\subsection*{Acknowledgements} 

I would like to thank L\'{a}szl\'{o} Erd\H{o}s
and Thomas Chen for numerous discussions
about the problems associated with deriving the crossing
estimate for the nearest neighbor interaction.
I am most grateful to Michael Loss for the instructive discussions 
enabling the use of the present, fairly general assumptions in the main
theorem.  I would also like to thank Herbert Spohn for the original
motivation of the problem and for several helpful remarks.
This work has been completed as part of the
Deutsche Forschungsgemeinschaft (DFG) project SP~181/19-1.

\section{Main results}
\label{sec:main}

Let us call a dispersion relation $\omega$
\defem{semi-dispersive}, if the integral over the modulus of its
resolvent diverges at most logarithmically, that is, if
there are $c_0\in \R_+$ and $n_0\in \N$ 
such that for all $0<\beta\le 1$,
and $\alpha\in \R$,
\begin{align}\label{eq:absresolvest}
&  \int_{\T^d} \rmd k\, 
\frac{1}{|\alpha -\omega(k)+\ci\beta|} \le c_0 \sabs{\ln \beta}^{n_0}.
\end{align}
We will be here mainly interested in real-analytic dispersion relations 
which have this property. 
We aim at proving (\ref{eq:RSEsuppr}), and thus we need to
consider the ``three-resolvent\footnote{This is to distinguish the estimate
  from the related integral involving four resolvent factors
  which was needed   in \cite{erdyau04} for the analysis going beyond the
  kinetic regime.}  
crossing integrals'' defined by 
\begin{align}\label{eq:defscint}
& \scint(\alpha,k_0,\beta) =
  \int_{(\T^d)^2} \rmd k_1\rmd k_2\, 
\frac{1}{|\alpha_1-\omega(k_1)+\ci\beta|
|\alpha_2-\omega(k_2)+\ci\beta|} 
\nonumber \\ & \qquad\times
\frac{1}{|\alpha_3-\omega(k_1-k_2+k_0)+\ci\beta|}
\end{align}
for $\alpha\in \R^3$, $k_0\in \T^d$ and $0<\beta\le 1$.
For any semi-dispersive $\omega$, we immediately obtain a bound for 
the integral by estimating the third factor trivially
by $1/\beta$, which yields
\begin{align}
\sup_{\alpha,k_0}\scint(\alpha,k_0,\beta) 
\le c_0^2 \sabs{\ln \beta}^{2 n_0} \beta^{-1}.
\end{align} 
We call this the \defem{basic estimate}.
We shall say that the \defem{dispersion relation suppresses 
crossings}, if it is possible to 
improve the basic estimate by some positive power of $\beta$, i.e., if 
there are constants $\gamma>0$, $c_1\in \R_+$, and $n_1\in\N$ such that
\begin{align}\label{eq2:crossingest}
& \sup_{\alpha,k_0}
 \scint(\alpha,k_0,\beta) \le c_1 \sabs{\ln \beta}^{n_1} \beta^{\gamma-1} .
\end{align}
We note that this implies, in particular, that 
$\sup_{\alpha,k_0} (\beta \scint(\alpha,k_0,\beta))\to 0$ when $\beta\to 0^+$.

The following collects the precise assumptions made here about $\omega$.
\begin{assumption}\label{th:DRass}
Let $d\ge 3$, and let $\omega:\R^d\to\R$ be real-analytic and 
$\Z^d$-periodic.  Define for all $s>0$,
\begin{align}\label{eq:deffom}
f_\omega(s) = 
\int_{\T^d}\! \rmd k\, \frac{1}{|\nabla \omega(k)|^3} \1(|\nabla 
\omega(k)|\ge s).
\end{align}
We assume that there are $p_0,c_0\ge 0$ such that for all $s>0$,
\begin{align}\label{eq:DRonly}
f_\omega(s) \le c_0 \sabs{\ln s}^{p_0}.
\end{align}
\end{assumption}
Since obviously $f_\omega(s)\le s^{-3}$, the assumption is really only
about the nature of the singularity of the integrand near 
the set of singular points of
$\omega$, i.e., about the behavior of $\omega$ near the points $k$ for
which $\nabla\omega(k)=0$.  

The first of the following theorems,
Theorem \ref{th:semidisp}, proves that
every such $\omega$ is semi-dispersive with $n_0=1$.  
This is the case, in particular,
for every real-analytic Morse function $\omega$ on $\T^d$, and we have
included a proof of this property 
in Appendix \ref{sec:Morse}.  In the assumptions,
for $d=3$ we then need to take $p_0=1$, otherwise $p_0=0$ suffices.
In the second theorem, Theorem \ref{th:thmain}, we present a simple geometric
classification of whether such a  dispersion relation 
suppresses crossings or not.
\begin{theorem}\label{th:semidisp}
Let Assumption \ref{th:DRass} be satisfied.  Then 
for every $0\le p\le 1$ there
is a constant $C_p$ with the following property: 
for all $\alpha\in \R$, and $0<\beta\le 1$, if $0\le p<1$,
  \begin{align} \label{eq:semidispsmallp}
 \int_{\T^d}\! \frac{\rmd k}{|\nabla\omega(k)|^p}
\frac{1}{|\alpha -\omega(k)+\ci\beta|} \le C_p \sabs{\ln \beta},
  \end{align}  
and, if $p=1$, 
  \begin{align} \label{eq:semidisp}
 \int_{\T^d}\! \frac{\rmd k}{|\nabla\omega(k)|}
\frac{1}{|\alpha -\omega(k)+\ci\beta|} \le C_1 \sabs{\ln \beta}^{p_0+2}.
  \end{align}  
\end{theorem}

\begin{theorem}\label{th:thmain}
Let Assumption \ref{th:DRass} be satisfied. 
Then $\omega$ suppresses crossings if and only if 
it is not a constant on any affine hyperplane.
\end{theorem}

Thus, we can now conclude that there is a large class of functions
for which the main Theorem in \cite{ls05} is satisfied:
\begin{corollary}
If $\omega:\T^3 \to \R$ is a Morse function, whose periodic extension to
$\R^3$ is real-analytic and the extension is not a constant on any
affine hyperplane, then it satisfies the assumptions (DR3) and
(DR4) of \cite{ls05}. 
\end{corollary}
The property called (DR3) was already shown in \cite{ls05} 
to be valid for Morse functions, we have included it in the Corollary only
to allow for easier use of the result.  With some effort,
it should now also be possible to generalize the results about the Anderson
model \cite{chen03} to more general dispersion relations.

\subsection{Notations}
\label{sec:defs}

We use the standard notations $S^d$ and $\T^d$ for the $d$-dimensional
unit sphere and the unit torus, respectively.  $S^d$ is the surface of the
unit ball in $\R^{d+1}$, with the topology and metric inherited from it,
and $\T^d$ is identified with the topological space $\R^d/\Z^d$.  We
denote the equivalence class mapping $\R^d\to \T^d$ by 
$[\cdot]$, and its inverse on $(-1/2,1/2]^d$ by $\tinj{\cdot}$.
The topology of the torus is then compatible with the
metric $\torusd$ defined by
$\torusd([y],[x])=\min_{n\in \Z^d}|y-x+n|=|\tinj{[y-x]}|$.
Let us also remark that, in general, 
we do not make a distinction between a periodic function
$f$ and its unique 
representative as a function on $\T^d$, defined by $[x]\mapsto f(x)$. 

The space dimension is denoted by $d$, and for any $r>0$, 
we denote the ball of radius $r$ in $\R^d$ by $B_r$.
In addition, we will reserve the notation $e_j$ to the $j$:th coordinate
vector of $\R^d$, i.e., $(e_j)_\nu = \delta_{j\nu}$, where $\delta$ denotes
the Kronecker delta.
An affine hyperplane $M\subset \R^d$ 
is a set for which there exists a vector $x_0\in \R^d$
such that $M-x_0$ is a hyperplane, i.e., a
$(d{-}1)$-dimensional subspace of $\R^d$.
Then there are a direction
$u\in S^{d-1}$ and $r_0\in \R$ such that with $x_0=r_0 u$,
$M=\defset{x\in \smash{\R^d}}{x\cdot u = r_0}=
\defset{x-(x\cdot u) u + x_0}{x\in \R^d}$.
We also denote the projection onto the
hyperplane orthogonal to $u$ by $Q_u$, when explicitly
\begin{align}
 Q_u x = x - (u\cdot x)u.
\end{align}

We use here the following standard shorthand notation
\begin{align}\label{eq:defsabs}
  \sabs{x} = \sqrt{1+x^2},
\end{align}
for $x\in \R$.  This will be the main tool for handling the various
power-law dependencies appearing later, and we have collected a few
basic properties of $\sabs{\cdot}$ into Appendix \ref{sec:sabsxprop}.
For any sufficiently many times differentiable 
function $f:X\to \C$, $X$ an open subset of $\R^{d}$, 
we employ the notations
\begin{align}\label{eq:deffnorms}
\norm{f}_{N} = \sup_{|\alpha|\le N} \norm{\partial^\alpha\! f}_\infty,
\quad\text{and}\quad
\norm{f}'_{N} = \sup_{0\le n\le N} \norm{D^n\! f}_\infty 
\end{align}
where, for a multi-index $\alpha$, $\partial^{\alpha}\! f$ is the corresponding
partial derivative of $f$, and, for a positive integer $n$, 
$D^n\! f|_x$ denotes the linear operator on 
$ \R^{d\times n}$
corresponding to the $n$:th derivative of $f$ at $x$.  Then
$ \norm{D^n\! f}_\infty = \sup_{x,|v_k|=1} 
\bigl|\prod_{k=1}^n (v_j\cdot\nabla)f(x)\bigr|$.
In particular,
$\norm{f}'_1 = \sup_x |\nabla f(x)|$, and 
$\norm{f}'_2 = \sup_x \norm{D^2 f(x)}$, where $D^2 f(x)$ is
the Hessian of $f$ at $x$ and the norm is its matrix norm.

Finally, $\1(P)$ denotes here a characteristic function of a statement $P$.
That is, it takes the value $1$, if $P$ is true, and $0$ otherwise.

\section{Counterexamples}
\label{sec:counterexamples}

\subsection{Proof of  ``only if'' in Theorem \ref{th:thmain}}

For this part of the proof, we do not need the dispersivity
properties following from Assumption \ref{th:DRass}, or the full
smoothness of the dispersion relation.  Instead of
the assumptions of Theorem \ref{th:thmain}, let us consider
in this subsection the following, more general, case:  
let $d\ge 2$ and assume that $\omega:\R^d\to \R$ is 
$\Z^d$-periodic and Lipschitz.  Let $C'$ denote
a Lipschitz constant of $\omega$, i.e., it is positive and
$|\omega(x')-\omega(x)|\le C' |x'-x|$ for all $x',x\in \R^d$.

To complete the ``only if'' part of Theorem \ref{th:thmain}, we assume
that there is an affine hyperplane $M\subset \R^d$
and $\alpha\in \R$ such that $\omega(x)=\alpha$ for all $x\in M$.
Then there are $u\in S^{d-1}$ and $r_0\in \R$ such that
$M=\defset{x\in \smash{\R^d}}{x\cdot u = r_0}=
\defset{x-(x\cdot u) u + x_0}{x\in \R^d}$, where $x_0=r_0 u$.
We shall prove that $\omega$ cannot suppress crossings by showing that
then there is $c>0$ such that for all $0<\beta\le 1$,
\begin{align}\label{eq:oisuffest}
I(\beta) = \scint((\alpha,\alpha,\alpha),[x_0],\beta) \ge \frac{c}{\beta}.
\end{align}
By the remark after (\ref{eq2:crossingest}), 
this suffices, as
then $\sup_{\alpha',k'_0} (\beta \scint(\alpha',k'_0,\beta))\ge c>0$.

We will derive the bound by considering the integral only over a certain 
neighborhood of $[M]\times [M]\subset \T^d\times \T^d$.  Let for any 
$\delta\ge 0$
\begin{align}\label{eq:defMp}
M'_\delta = \defset{[x]}{x\in \R^d,|x\cdot u-r_0|\le \delta}.
\end{align}
Then $M'_0 = [M]\subset M'_\beta$ and there is $C>0$ such that
$\int_{M'_\delta} \rmd k \ge C \delta$ for all $0\le \delta\le 1$.
If $k\in M'_\delta$, there is $x$ such that $k=[x]$ and
$|x\cdot u-r_0|\le \delta$.  Then $x'=x-(x\cdot u -r_0)u\in M$, and
\begin{align}
|\alpha - \omega(k)| = |\omega(x')-\omega(x)| \le
 C'|x\cdot u -r_0|  \le C'\delta .
\end{align}
Therefore, $|\alpha - \omega(k)+\ci \beta|  = 
\beta \sabs{(\alpha - \omega(k)) \beta^{-1}} 
\le \beta \sabs{C'\delta \beta^{-1}}$  for all $[x]\in M'_\delta$.
If $k_1$, $k_2\in M'_\beta$, then there are $x_1$, $x_2\in \R^d$ such that
$[x_j]=k_j$ and $|x_j\cdot u-r_0|\le \beta$.  Since
$(x_1-x_2+x_0)\cdot u - r_0 = x_1\cdot u-x_2\cdot u$, then
$[x_1-x_2+x_0]\in M'_{2\beta}$.  Therefore, for all $0<\beta\le 1$,
\begin{align}
& I(\beta) \ge 
  \int_{M'_\beta}\! \rmd k_1
  \int_{M'_\beta}\!\rmd k_2\, 
\frac{1}{\sabs{C'}^2 \sabs{2C'}} \beta^{-3}
 \ge \frac{C^2}{2 \sabs{C'}^3} \beta^{-1}.
\end{align}
This proves
(\ref{eq:oisuffest}), and finishes the proof of the ``only if''
part of Theorem \ref{th:thmain}.

\subsection{The first counterexample: NN-interaction in  $d=2$}

As the first counterexample, we consider the dispersion relation of the 
standard $2$-dimensional discrete Laplacian.  Although it does not satisfy
Assumption \ref{th:DRass}, as $d<3$ and $\omega$ is not analytic,
it is a standard example used in perturbative analysis of $2$-dimensional 
crystals.  We therefore find it worth the diversion to stress the special
nature of this dispersion relation, keeping in mind that the following
argument works quite generally for $2$-dimensional crystals with
translation invariant nearest neighbor (NN) -interactions.
See, for instance, sections 2.1 and 6 in
\cite{ls05} for more details on the subject.

Let $\omega:\R^2\to \R$ be defined by
\begin{align}
\omega(x) = \sqrt{2 - \cos(2\pi x_1) - \cos(2\pi x_2)}.
\end{align}
It is $\Z^2$-periodic and 
has a cusp singularity at every $x\in \Z^2$, but it is straightforward
to check that $\omega$ is nevertheless Lipschitz.  On the other hand,
if $x$ is any point on the affine hyperplane $x_1+x_2=\frac{1}{2}$, then
\begin{align}
  \omega(x)^2 = 2-\cos(2 \pi x_1)-\cos(\pi-2 \pi x_1) = 2.
\end{align}
Therefore, we can apply the previous proof, and conclude that the
dispersion relation $\omega$ does not suppress crossings.  The same is
naturally then true also for the dispersion relation $\omega^2$.

\subsection{Second example: A Morse function in $d=3$}

We want to provide also an example which satisfies Assumption
\ref{th:DRass} but is nevertheless a constant on a certain hyperplane,
to show that the extra condition in Theorem \ref{th:thmain} cannot be
dropped.  Define
\begin{align}
\omega(x) = 5 - \cos(2\pi x_1) 
\left(3+\cos(2\pi x_2)+\cos(2\pi x_3)\right)
\end{align}
which is $\Z^3$-periodic, real-analytic, and positive.
If we denote $s_j = \sin ( 2\pi x_j)$ and $c_j = \cos ( 2\pi x_j)$,
then
\begin{align}
& \frac{1}{2\pi} \nabla \omega(x) = (s_1(3+c_2+c_3), c_1 s_2, c_1 s_3).
\end{align}
Since $3+c_2+c_3 \ge 1$,
$\omega$ has $8$ critical points which are the points with
$s_j=0$ for all $j$, i.e., the points
$x_j\in\set{0,\frac{1}{2}}$ for $j=1,2,3$.  The Hessian is
\begin{align}
\frac{1}{(2\pi)^2} D^2 \omega(k) = \begin{pmatrix}
 c_1 (3+c_2+c_3) & -s_1 s_2 & -s_1 s_3 \\
 -s_1 s_2 & c_1 c_2 & 0 \\
 -s_1 s_3 & 0 & c_1 c_3 
\end{pmatrix}
\end{align}
and, since at all critical points $|c_j|=1$, if $x$ is a critical point
then $|\det D^2 \omega(x)|\ge (2\pi)^2>0$.
Therefore, $\omega$ is also a Morse function, and thus satisfies the
Assumption \ref{th:DRass}.  
On the other hand, $\omega(\pm  \frac{1}{4}, x_2, x_3) = 5$
for all $x_2$ and $x_3$, and thus $\omega$ is a constant, for
instance, on the hyperplane $x_1 = \frac{1}{4}$.  

As $\omega$ is positive, it is a dispersion relation of a certain 
classical harmonic crystal.  The corresponding elastic couplings of the
crystal can be obtained by taking the inverse Fourier
transform of $\omega^2$.  Since $\omega^2$ is a trigonometric polynomial, 
these elastic couplings correspond to a
translation invariant harmonic interaction which is mechanically stable
and has a finite range.  Therefore, this example shows that even quite simple
elastic couplings can lead to violation of the condition for suppression of
crossings.

\section{Main technical lemmas}
\label{sec:techlemmas}

We have collected in this section the technical material which will be
needed in the derivation of the main results.  We start with  
a few straightforward, but frequently applied, estimates.
In the second subsection we derive estimates for the asymptotics of
one-dimensional ``resolvent integrals''.  The final subsection contains 
a derivation of the parameterization of the level sets of $\omega$, and
most of it will be consumed by the more involved estimates about the higher
order curvature induced by the parameterization.

\subsection{Basic estimates}

For application of the following Lemmas, let us note that if 
$\omega$ satisfies the Assumption \ref{th:DRass}, then it is 
$\Z^d$-periodic and smooth, and thus $\norm{\omega}'_n<\infty$ for all $n$.   
\begin{lemma}\label{th:L1lemma}
Suppose $d$ and $\omega$
satisfy the Assumption \ref{th:DRass}.  Then for all $0<p<3$,
  \begin{align}\label{eq:L1bound}
     \int_{\T^d}\! \rmd k\, \frac{1}{|\nabla \omega(k)|^p} <\infty.
  \end{align}
\end{lemma}
\begin{proof}
Let $M=(\norm{\omega}'_1)^{3-p}$.  Then we can apply
a ``layer cake representation'' 
to the integral as in
\begin{align}
&     \int_{\T^d}\! \rmd k\, \frac{1}{|\nabla \omega(k)|^p} 
  =   \int_{\T^d}\! \rmd k\, \frac{1}{|\nabla \omega(k)|^3} 
 \int_0^{M}\! \rmd s\,  \1(|\nabla \omega(k)|^{3-p}\ge s) 
\nonumber \\ & \quad
=   \int_0^{M}\! \rmd s\, f_\omega\!\bigl(s^{\frac{1}{3-p}}\bigr) 
\le 
c_0 \Bigl\langle \frac{1}{3-p}\Bigr\rangle^{p_0}
\int_0^{M}\! \rmd s\, \sabs{\ln s}^{p_0} 
\end{align}
where we have used Fubini's theorem and
the general property $\sabs{a b}\le \sabs{a}\sabs{b}$.
By the change of variables to $y= -\ln s$, the remaining integral over $s$
is easily shown to be finite, 
which proves (\ref{eq:L1bound}).
\end{proof}

\begin{lemma}\label{th:nablaomdiff}
Let $a>0$ and $\omega:\R^d \to\R$, with $M_2=\norm{\omega}'_2<\infty$, 
be given.  Then for all $x,x_0\in \R^d$ with
$|x-x_0|\le \frac{a}{M_2} |\nabla \omega(x_0)|$,
\begin{align}\label{eq:nablaomdiff}
 |\nabla \omega(x)-\nabla \omega(x_0)| \le 
 a |\nabla \omega(x_0)|. 
\end{align}
\end{lemma}
\begin{proof}
Let $x$ and $x_0$ be such that 
$|x-x_0|\le \frac{a}{M_2} |\nabla \omega(x_0)|$. 
Choose an arbitrary $h\in \R^d$, when by Taylor formula and Schwarz
inequality,
\begin{align}
& |h\cdot(\nabla \omega(x)-\nabla \omega(x_0))|
\le \int_0^1 \!\rmd t \,\left|D^2\omega|_{x_0+t(x-x_0)}(h,x-x_0)\right|
\nonumber \\ & \quad
\le |h| \,|x-x_0|\, \norm{\omega}'_2 \le a |\nabla \omega(x_0)|\, |h|.
\end{align}
This proves (\ref{eq:nablaomdiff}).
\end{proof}

\begin{lemma}[argument shift] \label{th:idshift}
Let $\omega$ be such that $M_2=\norm{\omega}'_2<\infty$,
and assume that $s,p>0$ and $0<a<1$ are given.  Then for any 
$0<\lambda \le a s/M_2$, and $x,y\in \R^d$,
\begin{align}\label{eq:idshift}
&\frac{\1(|x-y|< \lambda)
\1(|\nabla \omega(y)|\ge s) }{|\nabla \omega(y)|^p} 
\nonumber \\ & \quad
\le (1+a)^p \frac{\1(|x-y|< \lambda)
\1(|\nabla \omega(x)|\ge (1-a)s) }{|\nabla \omega(x)|^p}.
\end{align}
\end{lemma}
\begin{proof}
Let us assume $|x-y|<\lambda$ and $|\nabla \omega(y)|\ge s$, otherwise the
bound in (\ref{eq:idshift}) is trivial.  
Since then $|x-y|< a |\nabla \omega(y)|/M_2$, we can apply
Lemma \ref{th:nablaomdiff} and triangle inequality, yielding
\begin{align}
|\,|\nabla \omega(x)|-|\nabla \omega(y)|\,| \le
|\nabla \omega(x)-\nabla \omega(y)| \le 
 a |\nabla \omega(y)|.
\end{align}
Therefore,
$|\nabla \omega(x)|\ge (1-a)|\nabla \omega(y)|\ge (1-a)s$,
and $ (1+a)|\nabla \omega(y)|\ge |\nabla \omega(x)|$,
which imply that (\ref{eq:idshift}) holds.
\end{proof}

\begin{lemma}\label{th:lnsint}
For any $p\ge 0$ and $0<\beta\le 1$,
\begin{align}\label{eq:lnsintbound}
  \int_{\beta}^1\!\rmd s\, \frac{\sabs{\ln s}^p}{s} \le 
\sabs{\ln \beta}^{p+1} .
\end{align}
\end{lemma}
\begin{proof}
Now $0\le -\ln s\le -\ln \beta$ for all $\beta\le s \le 1$.
Therefore,
\begin{align}
  \int_{\beta}^1\!\rmd s\, \frac{\sabs{\ln s}^p}{s}
 \le  \sabs{\ln \beta}^p \int_{\beta}^1 \frac{\rmd s}{s}
 = \sabs{\ln \beta}^p |\ln\beta| \le \sabs{\ln \beta}^{p+1} ,
\end{align}
proving (\ref{eq:lnsintbound}).
\end{proof}

\begin{lemma}\label{th:delbeta}
For any $\beta,\mu>0$, and $x,h\in \R$
such that $|h|\le 2 \mu \beta$,
\begin{align}\label{eq:delbineq}
 \frac{1}{|x+h+\ci \beta|} \le \frac{\mu+\sabs{\mu}}{|x+\ci \beta|} .
\end{align}
\end{lemma}
\begin{proof}
By the triangle inequality, $|x+h|^2 \ge (|x|-|h|)^2 $, and 
for any $0<\lambda<1$,
  \begin{align}
 &  |x+h+\ci \beta|^2 \ge x^2 - 2 |h| |x| + h^2 +\beta^2 
\nonumber \\ & \quad
   = (1-\lambda^2) ( x^2 + \beta^2 ) + 
   \bigl(\lambda |x| - \frac{1}{\lambda}|h|\bigr)^2 - 
\bigl(\frac{1}{\lambda^2}-1 \bigr) |h|^2
    +\lambda^2 \beta^2
\nonumber \\ & \quad
  \ge  (1-\lambda^2) ( x^2 + \beta^2 ) +   \beta^2 \bigl(
 (1-\frac{1}{\lambda^2} ) 4 \mu^2  +\lambda^2 \bigr).
  \end{align}
By choosing $\lambda^2 = 1-(\mu + \sabs{\mu})^{-2}$
the final term in the parenthesis vanishes.  Since then 
$1-\lambda^2 = (\mu + \sabs{\mu})^{-2}$, this proves (\ref{eq:delbineq}).
\end{proof}

\subsection{One-dimensional resolvent integrals}

We derive here the required estimates for one-dimensional ``resolvent''
integrals.  We start with polynomials, and
then extend these results to functions $f$ which are ``almost polynomial'' on
the integration interval in the sense that the $n_0$:th derivative of $f$
is non-vanishing on the whole interval for some order $n_0$.
The proof will be quite simple when $n_0=1$, and fairly involved when $n_0>1$.
Although we are not 
aware of a reference to a derivation of these estimates in the literature,
they could probably be pieced up from the known results.  We point out, in
particular, the similarity to Malgrange preparation theorem, see for instance
Section 7.5 of \cite{horm:PDE1}.  The main point of
reproducing the proofs in detail here is that we need to
have some control on how the various constants in the
estimates depend on the function $f$.

\begin{proposition}\label{th:polybound}
Let $n\ge 1$ and let $P_n(x)= \sum_{k=0}^n a_k x^k$, with $a_k\in \R$ and
$a_n\ne 0$.  If $n\ge 2$, then for all $\beta>0$, 
\begin{align}\label{eq:Pnint}
\int_{-\infty}^\infty \frac{ \rmd x}{|P_n(x)+\ci \beta|} \le 
\frac{2 (n+2)}{|a_n|^{1/n}}  \beta^{\frac{1}{n}-1} .
\end{align}
If $n=1$, then for $\beta,\lambda>0$, and $x_0\in\R$,
\begin{align}\label{eq:Pnintn1}
\int_{|x-x_0|\le \lambda} \frac{\rmd x}{|P_n(x)+\ci \beta|} \le 
\frac{6 \sabs{\ln \sabs{\lambda a_1}}}{|a_1|} \sabs{\ln \beta}.
\end{align}
\end{proposition}
\begin{proof}
Let first $n\ge 2$, and consider (\ref{eq:Pnint}).
Since $P_n$ is a polynomial of $n$:th degree, we can find
$z\in \C^n$ such that for all $x$, 
$P_n(x) = a_n \prod_{\ell=1}^{n} (x-z_\ell)$.
Fix then $x$, and let $\ell'$ be an integer such that 
$|x-z_\ell|\ge |x-z_{\ell'}|$ for all $\ell$.  Then,
$|x-z_\ell|\ge |x-\re z_{\ell'}|$, and
\begin{align}\label{eq:pneq1}
 \frac{1}{|P_n(x)+\ci \beta|} \le 
\frac{1}{\bigl||a_n| |x-\re z_{\ell'}|^n + \ci \beta\bigr|}
\le 
\sum_{\ell=1}^n 
\frac{1}{\bigl||a_n| |x-\re z_{\ell}|^n + \ci \beta\bigr|} .
\end{align}
For any $y\in \R$,
\begin{align}
\int_{-\infty}^\infty \frac{ \rmd x}{
  \bigl||a_n| |x-y|^n + \ci \beta\bigr|}
 = \frac{\beta^{\frac{1}{n}-1}}{|a_n|^{1/n}}
\int_{-\infty}^\infty \frac{\rmd x}{\sabs{x^n}},
\end{align}
where $\int_{-\infty}^\infty\! \rmd x\, \sabs{x^n}^{-1} \le
2 (1+\int_{1}^\infty\!\rmd x\, x^{-n})= 2n/(n-1)\le 2 (n+2)/n$, 
since $n\ge 2$.
Thus (\ref{eq:pneq1}) implies (\ref{eq:Pnint}).

Assume then $n=1$, when $P_n(x)=a_0+a_1 x$.  Changing variables to
$y=(a_0+a_1 x)/\beta$, we get
\begin{align}\label{eq:Pnintn1b}
\int_{|x-x_0|\le \lambda} \frac{\rmd x}{|P_n(x)+\ci \beta|} 
= \frac{1}{|a_1|} 
\int_{y_0-\lambda'}^{y_0+\lambda'} \frac{\rmd y}{|y+\ci|}
\end{align}
with $y_0=(a_0+a_1 x_0)/\beta$ and $\lambda'= |a_1|\lambda/\beta$.
By differentiation with respect to $y_0$, we find that the second integral
has a maximum at $y_0=0$.  Therefore,
\begin{align}\label{eq:Pnintn1c}
& \int_{|x-x_0|\le \lambda} \frac{\rmd x}{|P_n(x)+\ci \beta|} 
\le \frac{2}{|a_1|} 
\int_{0}^{\lambda'} \frac{\rmd y}{|y+\ci|}
\le \frac{2}{|a_1|}  (1+|\ln \lambda'|) 
\le \frac{2\sqrt{2}}{|a_1|} \sabs{\ln \lambda'}
\end{align}
If $\beta \le \lambda|a_1|$, then
$0 \le \ln \lambda' \le \ln \sabs{\lambda|a_1|}+|\ln \beta|$,
and, since  $2\sqrt{2} < 3$, (\ref{eq:Pnintn1c}) implies 
\begin{align}
& \int_{|x-x_0|\le \lambda} \frac{\rmd x}{|P_n(x)+\ci \beta|} 
\le \frac{3}{|a_1|} 2 \sabs{\ln \sabs{\lambda a_1}}\sabs{\ln \beta}
\end{align}
where we have used the properties of $\sabs{\cdot}$ given in Appendix
\ref{sec:sabsxprop}.  This proves (\ref{eq:Pnintn1}) for
$\beta \le \lambda|a_1|$.  If $\beta> \lambda|a_1|$, then we can
estimate trivially
\begin{align}
& \int_{|x-x_0|\le \lambda} \frac{\rmd x}{|P_n(x)+\ci \beta|} 
\le \frac{2\lambda}{\beta} < \frac{2}{|a_1|}
< \frac{6}{|a_1|} \sabs{\ln \sabs{\lambda a_1}}\sabs{\ln \beta},
\end{align}
which proves (\ref{eq:Pnintn1}) also for the remaining values of $\beta$.
\end{proof}

\begin{proposition}[$\bf n_0=1$]\label{th:genfboundn1}
Suppose $a,b\in \R$, with $a<b$.  Denote  $I=(a,b)$, and
assume $f\in C^{(1)}(I,\R)$ is such that
$|f'(x)|\ge \vep_0$ for some $\vep_0>0$ and all $x\in I$,
and that $m_0=\sup_{x\in I} |f(x)| < \infty$.
Then for all $\beta>0$ and $\alpha\in \R$,
\begin{align}\label{eq:fgenintn1}
\int_{a}^b \frac{ \rmd x}{|f(x)-\alpha+\ci \beta|} \le 
\frac{6 \sabs{\ln \sabs{m_0}}}{\vep_0} \sabs{\ln \beta}.
\end{align}
\end{proposition}
\begin{proof}
Since $f'$ is continuous, either $f'\ge \vep_0$ or 
$f'\le -\vep_0$, and we only need to prove the result in the first
case (applying it to $-f$ then proves the result in the second
case).   
Since $f'>0$, $f$ is strictly increasing. In addition,
$f(I)=(a',b')$, where 
$a'= \lim_{x\to a^{+}} f(x)$ and $b'= \lim_{x\to b^{-}} f(x)$ 
exist and are bounded by $m_0< \infty$.
Thus there is $g:f(I)\to I$,
$g=f^{-1}$, for which $g'(y) = 1/f'(g(y))\in (0,1/\vep_0]$.  Therefore,
\begin{align}
\int_{a}^b \frac{ \rmd x}{|f(x)-\alpha+\ci \beta|} = 
\int_{a'}^{b'} \!\rmd y \frac{g'(y)}{|y-\alpha+\ci \beta|} 
\le \frac{1}{\vep_0}
\int_{a'}^{b'} \frac{\rmd y}{|y-\alpha+\ci \beta|} .
\end{align}
By Lemma \ref{th:polybound}, this is bounded by
$6 \sabs{\ln \sabs{(b'-a')/2}} \sabs{\ln \beta}/\vep_0$.
However, as $|b'-a'|/2 \le m_0$, this bound implies also 
(\ref{eq:fgenintn1}).
\end{proof}

\begin{proposition}[$\bf n_0>1$]\label{th:genfbound}
Suppose $a,b\in \R$, with $a<b$, and  $n_0\ge 2$ are given.  
Denote  $I=(a,b)$, and assume $f\in C^{(n_0+1)}(I,\R)$
is such that $|f^{(n_0)}(x)|\ge n_0! \vep_0$ for some  $\vep_0>0$ and all 
$x\in I$, and that $m_0=\sup_{x\in I} |f^{(n_0+1)}(x)|/(n_0+1)! <\infty$.
Define $M=\max(m_0,1)$, $C_{n_0}=2^{n_0+1} (n_0+1)^{n_0}$, and
\begin{align}\label{eq:defepsp}
 \vep' = \frac{\vep_0}{M  C_{n_0}} >0 .
\end{align}
If $0< \beta\le (\vep')^{n_0+1}$, then
\begin{align}\label{eq:fgenint}
\int_{a}^b \frac{ \rmd x}{|f(x)+\ci \beta|} \le 
  C_{n_0} \left( \frac{b-a}{\vep_0} \beta^{\frac{1}{n_0+1}-1} +
 M \vep_0^{-\frac{1}{n_0}} \beta^{\frac{1}{n_0}-1} \right) .
\end{align}
\end{proposition}
\begin{proof}
We need to find the local 
minima of $|f|$, which coincide with the local minima of $f^2$.  
Since $f^{(n_0)}$ has no zeroes, $f^{(m)}$ has maximally $n_0-m$ zeroes
for $m\le n_0$.
Let $X$ be the union of the set of 
zeroes of $f$, of the zeroes of $f'$ and of the end-points $a$ and $b$,
when 
$|X|\le n_0 + n_0-1 + 2=2 n_0-1$.
Since $\rmd (f^2)/\rmd x=2 f f'$,
$X$ partitions $(a,b)$ into subintervals on which
$f^2$ --- and thus also $|f|$ ---
is strictly monotonic: if $a'<b'$ are such that 
$(a',b')\subset (a,b) \setminus X$, then $f^2$ is either strictly increasing
or decreasing on $[a',b']\cap (a,b)$.

Let us define $ \lambda = \beta^{\frac{1}{n_0+1}}$ when by assumption 
$0<\lambda\le \vep'$.  Suppose $x_0\in (a,b)$, and let 
$I=I(x_0)=\defset{x\in (a,b)}{|x-x_0|< \lambda}$. 
We claim that, if $x_0-\lambda> a$, 
then there is $x_0^-\in I$, $x_0^-<x_0$ such 
that $|f(x_0^-)|\ge M\vep' \lambda^{n_0}$,
and similarly, if $x_0+\lambda< b$, 
then there is $x_0^+\in I$, $x_0^+>x_0$ such 
that $|f(x_0^+)|\ge M\vep' \lambda^{n_0}$.
Consider the Taylor expansion of $f$ around $x_0$ to degree $n_0$,
\begin{align}
f(x) = \sum_{n=0}^{n_0} a_n (x-x_0)^n + R_{n_0}(x;x_0)
\quad\text{ where}\quad a_n= \frac{f^{(n)}(x_0)}{n!}.
\end{align}
For any $x$ there is a point $\xi$ between $x$ and $x_0$, such that 
the remainder is 
\begin{align}
R_{n_0}(x;x_0) = \frac{f^{(n_0+1)}(\xi)}{(n_0+1)!} (x-x_0)^{n_0+1},
\end{align}
implying that $|R_{n_0}| \le M \lambda^{n_0+1}$ on $I$.  On the other hand, 
since $|a_{n_0}|\ge \vep_0> 0$,
there is $z\in \C^{n_0}$ such that
\begin{align}
P_{n_0}(x;x_0) = \sum_{n=0}^{n_0} a_n (x-x_0)^n = a_{n_0} \prod_{j=1}^{n_0} (x-z_j).
\end{align}
Let $y_j = \re z_j$, when by $|x-z_j|\ge |x-y_j|$, we have for all $x\in I$,
\begin{align}\label{eq:PnRndiff}
|f(x)|\ge |P_{n_0}(x)| - |R_{n_0}(x)| \ge 
\vep_0 \prod_{j=1}^{n_0} |x-y_j| - M \lambda^{n_0+1}.
\end{align}
Consider the set $Y$ which consists of the endpoints of $I$ and of all
those $y_j$ which are in $I$.  Then $2\le |Y|\le n_0+2$.  The set
$[x_0-\lambda,x_0]\setminus Y \subset I$ 
consists of maximally $n_0+1$ intervals.  If $x_0-\lambda\ge a$, 
one of them
must be at least of length $\lambda/(n_0+1)$, and let $x_0^-$ be a
middle point of such an interval.  Then  $x_0^-<x_0$ and
$|x_0^- - y_j|\ge \frac{1}{2}\lambda/(n_0+1)$ for all $j$.  Therefore,
by (\ref{eq:PnRndiff}) and $\lambda\le \vep'$
\begin{align}
|f(x_0^-)| \ge 
\vep_0 \Bigl(\frac{\lambda}{2(n_0+1)}\Bigr)^{n_0}
 - M \lambda^{n_0+1} \ge 
(2 M \vep'-M \vep') \lambda^{n_0}
 = M\vep' \lambda^{n_0}.
\end{align}
If $x_0+\lambda\le b$, we can similarly find 
$x_0^+\in (x_0,x_0+\lambda]$ with 
$|f(x_0^+)|\ge M\vep' \lambda^{n_0}$.

For each $x_0 \in X$, we can thus find $x_0^{\pm}$ with the property that 
$x_0\in [x_0^-,x_0^+] \subset I(x_0)$, and either $x_0^{\pm}\in \set{a,b}$ 
or $|f(x_0^\pm)|\ge M\vep' \lambda^{n_0}$.  Let 
\begin{align}
X' = \defset{x_0 \in X}{|f(x_0)|< M\vep' \lambda^{n_0}}\qand
J=\bigcup_{x_0\in X'} (x_0^-,x_0^+).
\end{align}
We claim that if $x\in I\setminus J$, then 
$|f(x)| \ge M\vep' \lambda^{n_0}$. 

Suppose $x\in I\setminus J$.
It then belongs to an interval $I'$ whose endpoints
lie in the set $\cup_{x_0\in X'}\set{x_0^\pm}\cup\set{a,b}$.
Assume $x'$ is a local minimum point of $|f|$ on the closure of $I'$.
If $x'$ is not an endpoint of $I'$, 
it must be a critical point of $f^2$, and thus
$x'\in X$, when by construction, $|f(x')|\ge M\vep' \lambda^{n_0}$.
The same holds if $x'\in\set{a,b}\subset X$.  The only possibility left is
that $x'$ is one of the points $x_0^\pm$, when again by construction
$|f(x')|\ge M\vep' \lambda^{n_0}$.  This proves that 
$|f|\ge M\vep' \lambda^{n_0}$ on $I'$, in particular, also at $x$.

Therefore,
\begin{align}
& \int_{a}^b\! \rmd x\,\frac{1}{|f(x)+\ci \beta|} 
= \int_{I\setminus J}\! \rmd x\,\frac{1}{|f(x)+\ci \beta|}  +
\int_{J}\! \rmd x\,\frac{1}{|f(x)+\ci \beta|} 
\nonumber \\ & \quad
\le \frac{b-a}{M\vep' \lambda^{n_0}} +
 \sum_{x_0\in X'}
\int_{x_0^-}^{x_0^+}\! \rmd x\,\frac{1}{|f(x)+\ci \beta|} .
\end{align}
Consider one of the terms in the sum over $X'$, i.e., let $x_0\in X'$.
Denote $R(x)=R_{n_0}(x;x_0)$ and 
$P(x)=P_{n_0}(x;x_0)=a_{n_0} \prod_{j=1}^{n_0} (x-z_j)$.
Since $(x_0^-,x_0^+) \subset I(x_0)$, for all $x\in (x_0^-,x_0^+)$,
\begin{align}
|f(x)-P(x)| = |R(x)| \le M \lambda^{n_0+1}
= M\beta .
\end{align}
Therefore, by Lemma \ref{th:delbeta}, on the whole integration region
\begin{align}
\frac{1}{|f(x)+\ci \beta|} \le  \frac{\frac{1}{2}M+
 \sabs{\frac{1}{2}M}}{|P(x)+\ci \beta|} 
\le \frac{2 M}{|P(x)+\ci \beta|} 
\end{align}
to which we can apply Lemma \ref{th:polybound} with $|a_{n_0}|\ge \vep_0$.
Since $|X|\le 2 n_0-1$, the results proven so far can  be collected 
into the estimate
\begin{align}\label{eq:finalests}
& \int_{a}^b \frac{\rmd x}{|f(x)+\ci \beta|} 
\le \frac{b-a}{M\vep'} \beta^{\frac{1}{n_0+1}-1} + 
(2 n_0-1) 2 M \beta^{\frac{1}{n_0}-1}  \vep_0^{-\frac{1}{n_0}}
2 (n_0+2).
\end{align} 
To get the bound in  (\ref{eq:fgenint}), we  only need to use the
fact that, as $n_0\ge 2$, 
$C_{n_0}\ge 2^3 (n_0+1)^2\ge 2^2 (2 n_0-1)(n_0+2)$.
\end{proof}  

\subsection{Parameterization of the level sets}

The first of the results in this subsection 
states that, apart from the critical points,
there exists a local diffeomorphism which transforms the level sets of
$\omega$ into hyperplanes orthogonal to $e_1$.  Although
this is a straightforward consequence of the 
inverse mapping theorem, we need fairly detailed information
about the inverse function, and we have included also some
details of the proof here.  

{\em In all of the results in this subsection 
we assume that $d\ge 2$ and $\omega:\R^d \to \R$ is a smooth function such that
$\norm{\omega}'_n<\infty$  for all $n$\/}.  
In particular, this covers all dispersion relations satisfying Assumption
\ref{th:DRass}.
\begin{lemma}\label{th:standifth}
Let $x_0\in \R^d$ and $\lambda>0$ be such that 
$\nabla\omega(x_0)\ne 0$, and $\lambda\le \frac{1}{8}
\frac{|\nabla\omega(x_0)|}{\norm{\omega}'_2}$.  
Then there is an open set  
$U\subset \R^d$ and a diffeomorphism $\psi:B_{2\lambda}\to U$
with the following properties:
\begin{enumerate}
\item\label{it:diffe1} 
  $\psi(0)=x_0$ and $x_0+B_{\lambda}\subset \psi(B_{2\lambda})
  \subset x_0+B_{4\lambda}$.
\item\label{it:diffe2} For all $y$ with $|y|<2\lambda$,
\begin{align}
&\quad 
\omega(\psi(y)) = \omega(x_0)+ | \nabla\omega(x_0)| y_1,\qquad \text{and}
\label{eq:psiinvprop} \\
& |\nabla \omega(\psi(y)) -\nabla \omega(x_0)|  < 
 \frac{1}{2} |\nabla \omega(x_0)|.  \label{eq:nablaomdiff2}
\end{align}
\item\label{it:diffe3} Denote $A=D\psi(0)$
 and $u_0 = \nabla\omega(x_0)/|\nabla\omega(x_0)|$. 
Then $A$ is a rotation of $\R^d$ such that 
$u_0 = A e_1$.  In addition, $\frac{2}{3}\le |\det(D\psi)|\le 2$ on
$B_{2\lambda}$, and
\begin{align}\label{eq:Dpsieq}
\left. D\psi\right|_{y} A^T v =  v -   u_0 \left. 
\frac{\nabla\omega(x)\cdot v}{\nabla\omega(x)\cdot u_0} \right|_{x=\psi(y)} 
 \ \text{whenever}\quad  v\cdot u_0=0 .
\end{align}
\end{enumerate}
\end{lemma}
\begin{proof} 
Let us denote $U_a=x_0+B_{a 8\lambda}$, and 
define $f:\R^d\to \R^d$ by the formula
\begin{align}
f(x) = \frac{\nabla\omega(x)-\nabla\omega(x_0)}{|\nabla\omega(x_0)|}.
\end{align}
Then $f(x_0)=0$ and, by Lemma \ref{th:nablaomdiff}, $|f(x)| < a$
for all $x\in U_a$, $a>0$.  As before,
let $Q_{u_0}$ be the projection onto the subspace orthogonal to $u_0$,  
and let 
$O$ to be a rotation of $\R^d$ for which $O u_0 = e_1$; in particular,
$O^T=O^{-1}$ and $\det O =1$.
Define $\varphi:U_1 \to \R^d$ by
\begin{align}
\varphi(x) = \frac{\omega(x)-\omega(x_0)}{|\nabla\omega(x_0)|} e_1
 + O Q_{u_0}( x-x_0) .
\end{align}
Since $Q_{u_0} O^T e_1 =Q_{u_0} u_0=0$, then 
\begin{align}\label{eq:infphitcond}
 \varphi(x)_1 = \frac{\omega(x)-\omega(x_0)}{|\nabla\omega(x_0)|}.
\end{align}
By an explicit computation,
\begin{align}
D\varphi(x) = O + e_1 \otimes f(x) = O (\1 + u_0 \otimes f(x) ).
\end{align}
Since $O$ is orthogonal and $u_0 \otimes f(x)$ has rank one, the
determinant 
of $D\varphi(x)$ can be computed explicitly: with $u=f(x)$,
$\det D\varphi(x) = \det(\1 + e_1 \otimes (O u) ) 
= 1+(O u)_1 = 1+u_0\cdot u$
and thus for all $x\in U_a$,
\begin{align}\label{eq:detDphibounds}
  1-a < |\det D\varphi(x)| < 1+a.
\end{align}
Therefore,  $D\varphi(x)$ is invertible on $U_1$, and
by the inverse function theorem, $\varphi$ is
a local diffeomorphism on $U_1$.  Where we need to do the extra work here, 
is to show that we can find a neighborhood $U$ on which the inverse has
the properties stated in the Lemma.

Consider then the case $a=\frac{1}{2}$ in the above estimates.
Let $\phi(x)= O^T\!\varphi(x)-(x-x_0)$ for $x\in U_a$, when 
$\norm{D\phi(x)}<a$.
By the standard arguments used in the proof of the inverse function
theorem (see for instance the proof of Theorem 10.39 in \cite{rudin:fa}), 
it follows that $\varphi$ is one-to-one on $U_{a}$,
$B_{2\lambda}\subset \varphi(U_a)$, and 
$\psi=\left.\varphi^{-1}\right|_{B_{2\lambda}}$ is a diffeomorphism
from $B_{2\lambda}$ to an open set $U\subset U_a = x_0+B_{4\lambda}$.
Also, for all $y$,
\begin{align}\label{eq:Dpsieqraw}
D\psi(y) = D\varphi(\psi(y))^{-1}=
\Bigl(\1- \frac{1}{1+u_0\cdot u} u_0 \otimes u\Bigr)_{u=f(\psi(y))} O^T.
\end{align}

We now only need to check that $\psi$ has all the properties mentioned in the
Lemma. Since $\varphi(x_0)=0$, now $\psi(0)=x_0$ and we already proved 
$U\subset x_0+B_{4\lambda}$.  To complete item \ref{it:diffe1},
we need to prove that 
$U_{1/8}=x_0+B_{\lambda}\subset U$.  Since $U_{1/8}\subset U_{1/2}$, on
which $\varphi$ is one-to-one, it is enough to prove
$\varphi(U_{1/8})\subset B_{2\lambda}$.  This however holds now, since 
$\norm{D\varphi(x)}< 1+\frac{1}{8}$ for all  $x\in U_{1/8}$, and thus
$|\varphi(U_{1/8})|\le \frac{9}{8} \lambda<2\lambda$.
Of the two statements in item \ref{it:diffe2},
(\ref{eq:psiinvprop}) follows from (\ref{eq:infphitcond}) by
bijectivity of $\psi$, and, since $\psi(B_{2\lambda})\subset U_{1/2}$,
(\ref{eq:nablaomdiff2}) also holds.
For item \ref{it:diffe3}, we note that $A=D\psi(0)$ is equal to the
rotation $O^T$, 
and thus $A e_1=u_0$, and (\ref{eq:Dpsieqraw}) implies (\ref{eq:Dpsieq}).
Finally, by (\ref{eq:detDphibounds}) and 
$U\subset U_{\frac{1}{2}}$, we have 
$\frac{2}{3}\le |\det(D\psi(y))|\le 2$ for all $y$.
\end{proof}

\begin{corollary}\label{th:psicoroll}
Let $f:\R^d\to[0,\infty]$ be measurable. 
Then for any $x_0$, $\lambda$, and $\psi$ as in the previous Lemma,
  \begin{align}\label{eq:trickbound}
    \int_{|x-x_0|<\lambda} \!\rmd x\, f(x) \le
    2 \int_{|y|<2\lambda} \!\rmd y\,  f(\psi(y)) .
  \end{align}
\end{corollary}
\begin{proof}
By the properties of the diffeomorphism $\psi$ stated in the Lemma, 
  \begin{align}
&  \int_{|x-x_0|<\lambda} \!\rmd x\, f(x) \le
   \int_{\psi(B_{2\lambda})} \!\rmd x\, f(x)
 =\int_{|y|<2\lambda}\! \!\rmd y \left|\det(D\psi(y))\right| f(\psi(y))
  \end{align}
which is bounded by the right hand side of (\ref{eq:trickbound}).
\end{proof}

The final result in this section concerns the curvature induced on straight
lines by the ``level set diffeomorphism'' $\psi$.  In the following
Proposition we show that, if all derivatives of $\omega$ at $x_0$
in the direction
of the curve are small up to a certain order, then also the corresponding
``bending'' of the curve remains small up to the same order.  The
main difficulty in deriving these estimates lies in finding
sufficiently sharp estimates also when the
parameterization is nearly singular, i.e., when 
$|\nabla\omega(x_0)|\ll 1$.

\begin{proposition}\label{th:psiderivprop}
Let $\omega$, $x_0$ and $\lambda$ satisfy the assumptions of Lemma
\ref{th:standifth}, and let $\psi$, $A$, and $u_0$ be defined as in 
the conclusions of the Lemma.  Consider also some given
$|y|<2 \lambda$ and $v\in S^{d-1}$, with $v\cdot u_0=0$. 

Let $v'=A^T v$ and define 
\begin{align}
 \gamma(t;y,v) = \psi(y+t v')\qand 
\Gamma(t;y,v) = \gamma(t;y,v)-t v-\psi(y)
\end{align}
for all $t$ with $|y+t v'|<2\lambda$.
Then for any such $t$, and $n\ge 1$,
\begin{align}\label{eq:gnp1}
 \frac{1}{n!} \frac{\rmd^{n}}{\rmd t^{n}} \Gamma(t) = - g_n(t)  u_0
\end{align}
where
\begin{align}
g_n(t) = g_n(t;y,v) =
 \frac{1}{n!}\frac{\rmd^{n-1}}{\rmd t^{n-1}} g(\gamma(t;y,v))
\quad\text{with }\
g(x) = \frac{v\cdot\nabla\omega(x)}{ u_0\cdot\nabla\omega(x)},\
x\in \R^d.
\end{align}

Denote $M_n=\norm{\omega}'_n$, and $a_0=\max(1,8 M_2)$.
If $N\ge 2$, $0<\vep\le 1$, $\mu>0$ and $r_0>0$ are such that
$\mu\le (1+ 2^{N} + M_{N+1} 2^{2 N+1})^{-1}$,
$r_0\le \min(1,|\nabla\omega(x_0)|)$, 
$\lambda \le \vep (r_0\mu)^{N} a_0^{-1}$, and for all $2\le n<N$
  \begin{align}\label{eq:vcdotassump}
    \frac{1}{n!} \left|(v \cdot \nabla)^n \omega(x_0)\right| 
    \le \frac{1}{2} \vep (\mu r_0)^{N-n} ,
  \end{align}
then, with $C=1+\frac{M_N}{N!}$, 
\begin{align}\label{eq:DGnbounds}
\Bigl| \frac{1}{n!} \frac{\rmd^{n}}{\rmd t^{n}} \Gamma(t) \Bigr|
\le \begin{cases}
  \vep \mu^{N} r_0^{N-1}, & \text{ for }n=1\\
  2 \vep \mu^{N-n} r_0^{N-1-n}, & \text{ for }2\le n<N\\
 2 C r_0^{-1}, & \text{ for }n=N\\
 2 C \mu^{-1}  r_0^{-2}, & \text{ for }n=N+1
 \end{cases} .
\end{align}
\end{proposition}

The proof will be essentially a corollary of the following Lemma, whose
proof we will postpone until the end of this section:
\begin{lemma}\label{th:psiderivlemma}
Let the assumptions and definitions of the first paragraph of 
Proposition \ref{th:psiderivprop} be satisfied. 
Denote $M_n=\norm{\omega}'_n$, $a_0=\max(1,8 M_2)$, and assume
$0<r_0\le \min(1,|\nabla\omega(x_0)|)$ is given.  If 
$\lambda\le r_0 a_0^{-1}$, 
then all of the following results are valid:
\begin{enumerate}
\item\label{it:Gammabnd} $|\Gamma(t)|\le |t|<2\lambda$.
\item\label{it:defbarg}  
Let us define $\barg_n=\barg_n(x_0,v)$, by the following iterative 
procedure: 

Let $\barg_1 = 0$, $\barg_2 = \frac{1}{2}|\nabla\omega(x_0)|^{-1} (v\cdot
\nabla)^2\omega(x_0)$, 
and for $n>2$, define
\begin{align}\label{eq:gbardef}
  &  \barg_n = \frac{1}{|\nabla\omega(x_0)|}\Bigl[
  \frac{1}{n!} (v\cdot \nabla)^n\omega(x_0) 
  \nonumber \\ & \qquad +
  \sum_{k=2}^{n-1} \sum_{m\in \N_+^k} \1\Bigl(\sum_{j=1}^k m_j = n\Bigr)
  \prod_{j=1}^{k-1} \frac{m_j}{\sum_{j'=j}^{k} m_{j'}}
 \prod_{\substack{j=1 \\ m_j>1}}^k\!\barg_{m_j}
  \nonumber \\ & \qquad\quad  \times
    \left.(-u_0\cdot \nabla)^{k-\ell}  (v\cdot \nabla)^{\ell} \omega(x_0) 
    \right|_{\ell = \left|\defset{j}{m_j=1}\right|}
    \Bigr] .
\end{align}
Then $g_n(0;0,v)= \barg_n$ for all $n\ge 1$.
\item\label{it:bargbounds}
Suppose $0<\vep,\mu\le 1$
and $N\ge 2$ are such that for all $2\le n<N$, inequality
(\ref{eq:vcdotassump}) is satisfied. 
If $\mu\le 2^{-N} M_{N-1}^{-1}$, then for all $2\le m< N$,
\begin{align}\label{eq:bargbound}
  |\barg_m| \le \vep \mu^{N-m} r_0^{N-m-1} \le 1,
\end{align}
and, with $C=1+\frac{M_N}{N!}$ defined as in (\ref{eq:DGnbounds}),
\begin{align}\label{eq:bargNb}
 &  |\barg_N | \le C r_0^{-1}
\qand
  |\barg_{N+1} | 
 \le M_{N+1}
\left( 1 + 2^{N} C \right) r_0^{-2} .
\end{align}
Furthermore, if also
$b \ge 1+ 2^{N} + M_{N+1} 2^{2 N+1}$, then
for all $1\le n\le N$ and allowed $t$,
\begin{align}\label{eq:gammandiff}
  |g_n(t)-\barg_n| \le a_0 b^{n-1} \lambda r_0^{-n},
\end{align}
and
\begin{align}\label{eq:gammandiffN1}
  |g_{N+1}(t)-\barg_{N+1}| \le 5 C a_0 b^{N} \lambda r_0^{-N-2}.
\end{align}
\end{enumerate}
\end{lemma}
\begin{proofof}{Proposition \ref{th:psiderivprop}}
By Lemma \ref{th:standifth}, 
\begin{align}\label{eq:gt1d}
  \frac{\rmd}{\rmd t} \gamma(t) = v - g(\gamma(t)) u_0 
\end{align}
which implies (\ref{eq:gnp1}). 
For the results in the second paragraph, let us note 
that under the assumptions of the Proposition, 
we have $\lambda\le r_0/a_0$, so that items \ref{it:Gammabnd} and
\ref{it:defbarg} of Lemma \ref{th:psiderivlemma} are immediately
applicable. In addition, 
also $0<\mu\le 1$ with $\mu^{-1}\ge 2^N M_{N-1}$, so that if we define 
$b=\mu^{-1}$, then $b$ and $\mu$ are small enough for applying the
conclusions in item \ref{it:bargbounds}.  Therefore,
for $n=1$, we have $|\Gamma'(t)|= |g_1(t)|\le a_0 r_0^{-1} \lambda
\le \vep \mu^{N} r_0^{N-1}$, and if $2\le n<N$, then
\begin{align}\label{eq:DGn2N1}
& \Bigl| \frac{1}{n!} \frac{\rmd^{n}}{\rmd t^{n}} \Gamma(t) \Bigr|
= |g_n(t)| \le |g_n(t)-\barg_n| +|\barg_n| 
\nonumber \\ & \quad
\le a_0 b^{n-1} \lambda r_0^{-n} +
 \vep \mu^{N-n} r_0^{N-n-1}
\le 2 \vep \mu^{N-n} r_0^{N-n-1}  .
\end{align}
For $n=N$, we get similarly a bound 
$a_0 \mu^{1-N} \lambda r_0^{-N} + C r_0^{-1} \le 2 C r_0^{-1}$.
Finally, for $n=N+1$, we have
\begin{align}
& \Bigl| \frac{1}{n!} \frac{\rmd^{n}}{\rmd t^{n}} \Gamma(t) \Bigr|
\le 5 C a_0 \mu^{-N} \lambda r_0^{-N-2} + \mu^{-1} C r_0^{-2}
\le 2 C \mu^{-1} r_0^{-2}
\end{align}
where we have used $C'\le b C=\mu^{-1} C$ and $\mu\le \frac{1}{5}$.
This proves that all of the bounds given in 
(\ref{eq:DGnbounds}) are valid.
\end{proofof}

\begin{proofof}{Lemma \ref{th:psiderivlemma}}
For any $x=\gamma(t)$, we have in the definition of $g$
\begin{align}\label{eq:u0dotest}
 | u_0\cdot\nabla\omega(x)| =  
|\,|\nabla\omega(x_0)|+ u_0\cdot(\nabla\omega(x)-\nabla\omega(x_0))| \ge
\frac{1}{2}|\nabla\omega(x_0)|,
\end{align}
by (\ref{eq:nablaomdiff2}). Similarly,  $v\cdot u_0=0$ implies
\begin{align}\label{eq:vcdotest}
 | v\cdot\nabla\omega(x)| =  
| v\cdot(\nabla\omega(x)-\nabla\omega(x_0))|\le
|\nabla\omega(x)-\nabla\omega(x_0)|\le 4 \lambda M_2.
\end{align}
Therefore, (using the definition of $a_0$ and the assumption 
made on $\lambda$)
\begin{align}\label{eq:g1bound2}
|g_1(t)| \le a_0 \lambda r_0^{-1} \le 1,
\end{align}
which implies
$|\Gamma'(t)| \le 1$. Since $\Gamma(0)=0$, item \ref{it:Gammabnd} holds now.

Consider then item \ref{it:defbarg}.
In (\ref{eq:gbardef}), the sum over $m_j$ is restricted by $k\ge 2$
so that always $m_j\le n-m_1\le n-1$. Thus the right hand side depends
only on $\barg_m$ with $2\le m\le n-1$, and the sequence $\barg_n$ is
thus uniquely determined from $\barg_2$ and it only depends on $x_0$ and
$v$ (and naturally also on $\omega$).   To complete the proof of the item, 
we need to show that $\barg_n=g_n(0;0,v)$.  We do this by induction: 
Since $g_1(0;0,v)=0=\barg_1$, this holds for $n=1$. 
Let us assume that
the result is true for $1\le m<n$.
By Lemma \ref{th:standifth}, we have for all $t$, 
$\omega(\gamma(t))= \omega(x_0)+|\nabla\omega(x_0)| y_1$, which is
independent of $t$.  By Lemma
\ref{th:compdiff} the $n$:th derivative of 
$\omega\circ \gamma$, which is zero, can be expressed
in terms of differentials of $\gamma$.  
We separate the $k=1$ term in the resulting sum, yielding
\begin{align}\label{eq:gammaneq}
& -\frac{\gamma^{(n)}(t)}{n!}\cdot \nabla \omega(\gamma(t))
 = \sum_{k=2}^n \sum_{m\in \N_+^k} \1\Bigl(\sum_{j=1}^k m_j = n\Bigr)
 \prod_{j=1}^{k-1} \frac{m_j}{\sum_{j'=j}^{k} m_{j'}}
\nonumber \\ & \qquad
\times \prod_{j=1}^k\! \left.\left[ 
 \smash{\frac{1}{m_j!}}\gamma^{(m_j)}(t) \cdot \nabla \right] \!
\omega \right|_{\gamma(t)} .
\end{align}
At $t=0$ and $y=0$, $\gamma(t)=x_0$ and $\gamma^{(1)}(t)=v$, and
the left hand side evaluates to $g_n(0;0,v) |\nabla\omega(x_0)|$.
Since the induction assumption can be applied to all derivatives of
$\gamma$ in the right hand side, we find that it evaluates to 
right hand side of (\ref{eq:gbardef}) times $|\nabla\omega(x_0)|$.
This completes the induction step and proves $g_n(0;0,v)= \barg_n$.

We next prove the statements in item \ref{it:bargbounds}.
If $N=2$, then (\ref{eq:bargbound}) is vacuously true, 
and (\ref{eq:bargNb}) holds by an explicit
computation.  Consider then $N>2$, when again an explicit
computation proves that
(\ref{eq:bargbound}) holds for $n=2$.  We will prove its validity for higher
values of $n$ by induction.   Let us thus assume that $2\le n \le N$ is 
given and that
(\ref{eq:bargbound}) is valid for all $2\le m <n$.
Suppose $2\le k\le n-1$ and 
$\sum_{j=1}^k m_j =n$, and let $\ell=\left|\defset{j}{m_j=1}\right|$.
Then $0\le \ell\le k-1$, 
and $\sum_{j,m_j>1} (1-m_j) = \sum_{j} (1-m_j) = k-n$.  Therefore,
since $0<\vep, \mu, r_0\le 1$, and $k\ge 2$,
\begin{align}\label{eq:prodbarg}
& \Bigl| \prod_{\substack{j=1, m_j>1}}^k\!\barg_{m_j} \Bigr|
\le \left( \vep \mu^{N-1} r_0^{N-2} \right)^{k-\ell}
 (\mu r_0)^{\sum_{j,m_j>1} (1-m_j)}
 \nonumber \\ & \quad 
\le \vep \mu^{N-1+k-n} r_0^{N-2+k-n}
 \le \vep \mu^{N+1-n} r_0^{N-n}.
\end{align}
Using this estimate in (\ref{eq:gbardef}) yields
\begin{align}\label{eq:barg22}
 &  \Bigl|\barg_n - \frac{1}{|\nabla\omega(x_0)|}
  \frac{1}{n!} (v\cdot \nabla)^n\omega(x_0) \Bigr| 
  \nonumber \\ &\quad  
\le \frac{1}{r_0} 
\sum_{k=2}^{n-1} \sum_{m\in \N_+^k} \1\Bigl(\sum_{j=1}^k m_j = n\Bigr)
\vep \mu^{N+1-n} r_0^{N-n} M_{k}
  \nonumber \\ &\quad  
\le \vep \mu^{N-n} r_0^{N-n-1} \mu M_{N-1} 2^{n-1}
\le \frac{1}{2} \vep \mu^{N-n} r_0^{N-n-1}
\end{align}
where we have applied the assumption made on $\mu$, and
the equality (provable, e.g., by induction, or
by a combinatorial argument for all $n\ge 1$ and $1\le k\le n$)
\begin{align}
\sum_{m\in \N_+^k} \1\Bigl(\sum_{j=1}^k m_j = n\Bigr) = \binom{n-1}{k-1}.
\end{align}
If $n<N$, we can then apply the assumption (\ref{eq:vcdotassump})
to (\ref{eq:barg22}) and obtain the bound
\begin{align}
 &  |\barg_n | 
 \le  \frac{1}{r_0}  \frac{1}{2} \vep (\mu r_0)^{N-n} +
 \frac{1}{2} \vep \mu^{N-n} r_0^{N-n-1}
 \le   \vep \mu^{N-n} r_0^{N-n-1} .
\end{align}
This completes the induction step and proves (\ref{eq:bargbound})
for $2\le m<N$.  However, then (\ref{eq:barg22}) is valid also for $n=N$, 
and thus also
\begin{align}
 &  |\barg_N | 
 \le \frac{1}{r_0} \frac{M_N}{N!} +
   \frac{1}{2} \vep  r_0^{-1} 
 \le \Bigl( \frac{M_N}{N!} + \frac{1}{2}\Bigr) r_0^{-1} .
\end{align}
Finally, then for any $2\le k\le N$ and $m\in \N_+^{k}$
such that $\sum_j m_j = N+1$,
\begin{align}\label{eq:prodbarg2}
& \Bigl| \prod_{\substack{j=1, m_j>1}}^k\!\barg_{m_j} \Bigr|
\le  C r_0^{-1}.
\end{align}
To see this, note that $|\barg_N|$ can appear in the product only once, and
the other factors are always less than one.
Therefore, as in (\ref{eq:barg22}), we find
\begin{align}
 &  |\barg_{N+1} | 
 \le \frac{1}{r_0} \frac{M_{N+1}}{(N+1)!} +  \frac{1}{r_0}
  \sum_{k=2}^N \binom{N}{k-1} M_N C r_0^{-1}
\end{align}
which yields the bound in (\ref{eq:bargNb}).

We still need to prove (\ref{eq:gammandiff}).  By (\ref{eq:g1bound2}), it
holds for $n=1$, so let us assume that $n\ge 2$.
By (\ref{eq:gnp1}), the left hand side of (\ref{eq:gammaneq}) is then 
equal to 
$g_n(t) u_0 \cdot \nabla \omega(\gamma(t))$, implying
\begin{align}
& \Bigl| g_n(t) |\nabla\omega(x_0)| + 
 \frac{\gamma^{(n)}(t)}{n!}\cdot \nabla \omega(\gamma(t))
 \Bigr| \le  \left| g_n(t) \right| |\nabla\omega(x_0)-
\nabla \omega(\gamma(t))|
\nonumber \\ & \quad
 \le  \left| g_n(t) \right| M_2 4 \lambda
 \le  M_2 4 \lambda\left| \barg_n \right| 
 + \frac{1}{2}|\nabla\omega(x_0)| \left| g_n(t) - \barg_n\right|.
\end{align}
Therefore, by employing the triangle inequality to change $g_n(t)$ to
$\barg_n$ in the leftmost expression, we find that
\begin{align}\label{eq:gnbargdiff}
&| g_n(t) -\barg_n | \le 
 \frac{8 M_2}{r_0} \lambda \left| \barg_n \right|  + \frac{2}{r_0}
  \Bigl||\nabla\omega(x_0)|\barg_n+
  \frac{\gamma^{(n)}(t)}{n!}\cdot \nabla \omega(\gamma(t))
 \Bigr|  .
\end{align}

We next need to bound the right hand side of (\ref{eq:gammaneq}) minus 
$|\nabla\omega(x_0)| \barg_n$. Using the definition of $\barg_n$, we
get a bound
\begin{align}\label{eq:gammandiffb}
&\sum_{k=2}^n \sum_{m\in \N_+^k} \1\Bigl(\sum_{j=1}^k m_j = n\Bigr)
 \prod_{j=1}^{k-1} \frac{m_j}{\sum_{j'=j}^{k} m_{j'}}
\nonumber \\ & \quad\times 
\biggl| \prod_{j=1}^k\! \left.\left[ 
 \smash{\frac{1}{m_j!}}\gamma^{(m_j)}(t) \cdot \nabla \right] \!
\omega \right|_{\gamma(t)} 
-  \prod_{\substack{j=1, \\ m_j>1}}^k\!
    (-\barg_{m_j} u_0\cdot \nabla)
 \prod_{\substack{j=1, \\ m_j=1}}^k\! (v\cdot \nabla) \omega(x_0) 
\biggr| .
\end{align}
Here the absolute value needs to be bounded, and we do this in  
two steps: first we
shift $\gamma'(t)$ to $v$ and higher derivatives to $\barg$
by using the induction assumption 
and then we shift the valuation point from $\gamma(t)$ to $x_0$.  

To illustrate this, let us perform the estimates first
for the case $k=n$, when the induction assumption is not needed, and we can
therefore apply the result for any $n$.
Then the absolute value is explicitly
\begin{align}\label{eq:gammandiffb3}
& \Bigl| \left[(v-g_1(t) u_0) \cdot \nabla \right]^n \!
\omega(\gamma(t))
-  (v\cdot \nabla)^n \omega(x_0) \Bigr| 
\nonumber \\ & \quad
\le \Bigl| \left[(v-g_1(t) u_0) \cdot \nabla \right]^n \!
\omega(\gamma(t))
-  (v\cdot \nabla)^n \omega(\gamma(t)) \Bigr| 
\nonumber \\ & \qquad
+ \Bigl|  (v\cdot \nabla)^n \omega(\gamma(t))
-  (v\cdot \nabla)^n \omega(x_0) \Bigr| 
\nonumber \\ & \quad
\le \sum_{j=1}^n \binom{n}{j} |g_1(t)|^j M_n
+ M_{n+1} |\gamma(t)-x_0|
\nonumber \\ & \quad
\le M_{n+1}
\Bigr[\sum_{j=1}^n \binom{n}{j} |g_1(t)| + 4 \lambda \Bigl]
\le a_0\lambda r_0^{-1} M_{n+1} 2^{n+1}
\end{align}
where we have used the Leibniz rule.
But now (\ref{eq:gnbargdiff}) implies that for $n=2$, 
\begin{align}
| g_2(t) -\barg_2 | \le a_0\lambda r_0^{-1} |\barg_2|
 + 2 a_0 \lambda r_0^{-2} M_{3} 2^{3}.
\end{align}
If $N=2$, (\ref{eq:bargNb}) implies then that 
\begin{align}
| g_2(t) -\barg_2 | \le a_0\lambda r_0^{-2} \bigl(
 \frac{1}{2} ( 1+M_2 ) +
M_{3} 2^{4}\bigr ) \le a_0\lambda r_0^{-2} b|_{N=2}. 
\end{align}
If $N>2$, by (\ref{eq:bargbound}) $|\barg_2|\le 1$, and thus
\begin{align}
| g_2(t) -\barg_2 | \le a_0\lambda r_0^{-2} (1+ M_{3} 2^{3})
\le a_0 \lambda r_0^{-2}  b.
\end{align}
This proves that (\ref{eq:gammandiff}) holds always for $n=2$.

Let us then make the induction assumption that $2<n\le N$
and (\ref{eq:gammandiff}) holds for all $2\le m<n$.  The case $k=n$ has
already been treated above, so let us assume $k<n$.
We begin by estimating the result from the second step.
Let $\ell =\left|\defset{j}{m_j=1}\right|$, which now satisfies $\ell < k$.
Since $k>1$, we also have $m_j\le n-1$ for all $j$, and by
(\ref{eq:bargbound}), now 
$\prod_{\substack{j=1, \\ m_j>1}}^k\! |\barg_{m_j}|\le 1$.
Therefore,
\begin{align}\label{eq:gminusx0}
& \Bigl|\prod_{\substack{j=1, \\ m_j>1}}^k\!
    (-\barg_{m_j} u_0\cdot \nabla)(v\cdot \nabla)^\ell \omega(\gamma(t)) 
-\prod_{\substack{j=1, \\ m_j>1}}^k\!
    (-\barg_{m_j} u_0\cdot \nabla)(v\cdot \nabla)^\ell \omega(x_0) 
\Bigr| 
\nonumber \\ & \quad
 \le |\gamma(t)-x_0| M_{k+1}
\prod_{\substack{j=1, \\ m_j>1}}^k\! |\barg_{m_j}| 
 \le 4  \lambda M_{n} \le b \lambda \le a_0 b^{n-2} \lambda .
\end{align}

To estimate the result from the first step, 
let $I_k=\set{1,2,\ldots,k}$. 
Using the commutativity of the partial derivatives,
the result can be bounded by
\begin{align}\label{eq:firststep}
& \sum_{\substack{I\subset I_k \\ I\ne \emptyset}} 
  \Bigl|  \prod_{j\in I}
    (g_{m_j}(t)-\barg_{m_j}) 
    \prod_{\substack{ j\not\in I,\\ m_j>1 }} 
    (-\barg_{m_j} u_0\cdot \nabla)
    \prod_{\substack{ j\not\in I,\\ m_j=1 }} 
    (-v\cdot \nabla)
    (-u_0\cdot \nabla)^{|I|}\omega(\gamma(t)) \Bigr|
\nonumber \\ & \quad 
\le M_k 
 \sum_{\substack{I\subset I_k \\ I\ne \emptyset}} 
 \prod_{\substack{ j\in I }}
   |g_{m_j}(t)-\barg_{m_j}|
\le M_k 
 \sum_{\substack{I\subset I_k \\ I\ne \emptyset}} 
 (a_0\lambda r_0^{-1})^{|I|}
 \prod_{\substack{ j\in I }} (b r_0^{-1})^{m_j-1}
\nonumber \\ & \quad 
\le M_{n-1} 
 \sum_{\substack{I\subset I_k \\ I\ne \emptyset}} 
 a_0\lambda r_0^{-1} (b r_0^{-1})^{n-2}
\le M_{n-1} 2^k
 a_0\lambda r_0^{1-n} b^{n-2}.
\end{align}
where we have applied
\begin{align}
\sum_{j\in I} (m_j-1) \le \sum_{j=1}^k (m_j-1) = n-k \le n-2,
\end{align}
and, as $I\ne \emptyset$ and $a \lambda\le r_0$, we have also
$(a \lambda/r_0)^{|I|}\le a \lambda/r_0$.
Combining the above estimates, we then have obtained
the following bound for (\ref{eq:gammandiffb}):
\begin{align}\label{eq:gmdb2}
& \sum_{k=2}^n \binom{n-1}{k-1} a_0 b^{n-2} \lambda r_0^{1-n}
\left( 1 + M_n 2^{n} \right)
\nonumber \\ & \quad
\le  a_0 b^{n-2} \lambda r_0^{1-n} 2^{n-1}
\left( 1 + M_n 2^{n} \right).
\end{align}
Therefore, (\ref{eq:gnbargdiff}) now implies that for any $n< N$,
\begin{align}\label{eq:gnbargdiff2}
&| g_n(t) -\barg_n | \le 
 \frac{a_0}{r_0} \lambda +
 a_0 b^{n-2} \lambda r_0^{-n} 2^{n}
\left( 1 + M_n 2^n \right)
\nonumber \\ & \quad
\le \frac{1+ 2^{n} + M_n 2^{2 n}}{b} 
a_0 b^{n-1} \lambda r_0^{-n} 
\le a_0 b^{n-1} \lambda r_0^{-n} 
\end{align}
by our choice of $b$.  This completes the induction step and proves
that (\ref{eq:gammandiff}) is valid for all $2\le n<N$.  However, 
then we can still use the bound
(\ref{eq:gmdb2}), together with (\ref{eq:bargNb}), 
in (\ref{eq:gnbargdiff}) which shows that
\begin{align}
  |g_N(t)-\barg_N| \le \frac{ M_N + 1
+ 2^{N} + M_N 2^{2 N}}{b} a_0 b^{N-1} \lambda r_0^{-N}
\le a_0 b^{N-1} \lambda r_0^{-N} .
\end{align}
This proves that $b$ is large enough for (\ref{eq:gammandiff}) 
to hold also for $n=N$.  For $n=N+1$, we repeat the above steps 
using (\ref{eq:prodbarg2}), and the fact that (\ref{eq:gammandiff}) 
holds also for $n=N$.  Then the left hand sides of equations
(\ref{eq:gminusx0}) and (\ref{eq:firststep}) can be
bounded by $4 \lambda M_{N+1} C r_0^{-1}$ and
$2^{N+1}M_{N+1} C r_0^{-1-N}\lambda a_0 b^{N-1}$,
respectively. This yields a bound 
$2^{2 N+2}M_{N+1} a_0 b^{N-1} C r_0^{-1-N}\lambda$
for (\ref{eq:gammandiffb}). Then using
the bound for $|\barg_{N+1}|$ given 
in (\ref{eq:gnbargdiff}) proves that
\begin{align}\label{eq:gnbargdiff3}
&| g_{N+1}(t) -\barg_{N+1} | \le 
 \frac{a_0}{r_0} \lambda C' r_0^{-2} +
C r_0^{-2-N} \lambda a_0 b^{N-1} 2^{2 N+3} M_{N+1}
\end{align}
where $C'=M_{N+1} \left( 1 + 2^{N} C \right)\le b C$.
Finally, using $2^{2 N+1} M_{N+1} \le b$,
proves (\ref{eq:gammandiffN1}).  
\end{proofof}

\section{Semi-dispersivity (Proof of Theorem \ref{th:semidisp})}
\label{sec:proofofsemi}

Let $0\le p\le 1$, $\alpha\in \R$, and $0<\beta\le 1$ be arbitrary, and
denote  $M_n=\norm{\omega}'_n$ for all $n$. Let us
define further $q=1+\frac{1}{2}(1-p)$, so that, if $p=1$, also $q=1$, and otherwise
$q+p+1<3$.  We then apply the layer cake representation as
\begin{align}
&  \int_{\T^d}\! \frac{\rmd k}{|\nabla\omega(k)|^p}
\frac{1}{|\alpha -\omega(k)+\ci\beta|}
= \int_0^{\Mf^q}\! \rmd s\,
\int_{\T^d}\!
\frac{ \rmd k}{|\alpha -\omega(k)+\ci\beta|} 
\frac{\1(|\nabla \omega(k)|\ge s^{\frac{1}{q}}) }{|\nabla \omega(k)|^{p+q}}
\nonumber \\ & \quad
\le \int_0^{\beta}\! \frac{ \rmd s}{\beta}
\int_{\T^d}\!\frac{\rmd k}{|\nabla \omega(k)|^{p+q}}
+  \int_\beta^{\Mf^q}\!\! \rmd s\!
\int_{\T^d}\!
\frac{ \rmd k}{|\alpha -\omega(k)+\ci\beta|} 
\frac{\1(|\nabla \omega(k)|\ge s^{\frac{1}{q}}) }{|\nabla \omega(k)|^{p+q}} .
\end{align}
Since $p+q\le 2$, the first term is bounded by a
$\beta$-independent constant by Lemma \ref{th:L1lemma}.
To analyze the second term,
let us define the following cut-off function $G:\R^d\times (0,1/2]\to \R$,
\begin{align}
 G(x,\lambda) = \frac{N_d}{\lambda^d} \1(|x|< \lambda)
\end{align}
where $N_d=d/|S^{d-1}|$ is a normalization constant such that
$\int_{\R^d}\! \rmd x\, G(x,\lambda) = 1$  for all $\lambda$. 
We have restricted the range of $\lambda$ in the above manner so that 
for all $k\in \T^d$ and $\lambda$ we still have
$\int_{\T^d}\! \rmd x\, G(\tinj{x-k},\lambda) = 1$ (we recall the
definition of $\tinj{\cdot}$ in section \ref{sec:defs}).

By choosing
$\lambda=\lambda(s)=\min(\frac{1}{4},s^{\frac{1}{q}}/(9 M_2))$, 
we then find
\begin{align}\label{eq:1dxkint}
& \int_{\T^d}\!
\frac{ \rmd k}{|\alpha -\omega(k)+\ci\beta|} 
\frac{\1(|\nabla \omega(k)|\ge s^{\frac{1}{q}}) }{|\nabla \omega(k)|^{p+q}} 
\nonumber \\ & \quad
 = \int_{\T^d}\! \rmd x\, \int_{\T^d}\! \rmd k\,
\frac{G(\tinj{x-k},\lambda)}{|\alpha -\omega(k)+\ci\beta|} 
\frac{\1(|\nabla \omega(k)|\ge s^{\frac{1}{q}}) }{|\nabla \omega(k)|^{p+q}}.
\end{align}
Applying Lemma \ref{th:idshift} with $a=1/9$
shows that this is bounded by 
\begin{align}\label{eq:1dxkintb}
(1+a)^q \int_{\T^d}\! \rmd x
\frac{\1(|\nabla \omega(x)|\ge (1-a) s^{\frac{1}{q}})}{|\nabla \omega(x)|^{p+q}}
 \frac{N_d}{\lambda^d} \int_{\R^d}\! \rmd k\,
\frac{\1(|\tinj{x}-k|< \lambda)}{|\alpha -\omega(k)+\ci\beta|}  .
\end{align}
Let  $x_0=\tinj{x}$. Then inside the integral
$\lambda\le \frac{|\nabla \omega(x_0)|}{8 M_2}$
since $9 (1-a)= 8$.
Therefore, Lemma \ref{th:standifth} yields a diffeomorphism $\psi$, such
that we can apply Corollary \ref{th:psicoroll}.  This shows that
\begin{align}
& \frac{N_d}{\lambda^d} \int_{|x_0-k|< \lambda}\!
\frac{ \rmd k}{|\alpha -\omega(k)+\ci\beta|} 
\le \frac{2 N_d}{\lambda^d} \int_{|y|<2\lambda}\!
\frac{\rmd y}{|\alpha -\omega(\psi(y))+\ci\beta|} 
\nonumber \\ & \quad
\le \frac{2^d N_d}{\lambda} \frac{|S^{d-2}|}{d-1}
 \int_{-2\lambda}^{2\lambda}\! 
\frac{\rmd y_1}{|\alpha -\omega(x_0)-|\nabla\omega(x_0)|y_1+\ci\beta|} 
\nonumber \\ & \quad
\le \frac{2^d N_d}{N_{d-1}}
\frac{6 \sabs{\ln \sabs{2 \lambda |\nabla\omega(x_0)|}}}{
  \lambda|\nabla\omega(x_0)|} \sabs{\ln \beta}
\le \frac{2^d N_d}{N_{d-1}}
\frac{6 \sabs{\ln \sabs{M_1}}}{
  \lambda|\nabla\omega(x_0)|} \sabs{\ln \beta}
\end{align}
where we have applied Lemma \ref{th:polybound},
and the properties of $\sabs{\cdot}$ given in Appendix \ref{sec:sabsxprop}
together with $0\le 2 \lambda |\nabla\omega(x_0)|\le M_1$.
Combining this with (\ref{eq:1dxkintb}) and (\ref{eq:1dxkint}), we have
proven that there is  a constant $c'\ge 1$, which
depends only on $M_1=\norm{\omega}'_1$, such that
\begin{align}
& \int_{\T^d}\!\frac{ \rmd k}{|\alpha -\omega(k)+\ci\beta|} 
\frac{\1(|\nabla \omega(k)|\ge s^{\frac{1}{q}}) }{|\nabla \omega(k)|^{p+q}} 
\nonumber \\ & \quad
\le \frac{(1+a)^q c'}{\lambda(s)} \sabs{\ln \beta} \int_{\T^d}\! \rmd x
\frac{\1(|\nabla \omega(x)|\ge (1-a) 
  s^{\frac{1}{q}})}{|\nabla \omega(x)|^{p+q+1}}.
\end{align}

If $p<1$, then $p+q+1<3$ and, by Lemma \ref{th:L1lemma}, the remaining
integral over $x$
can be bounded by a constant independent of $s$.  After this, the
integral over $s$ only yields a factor
\begin{align}
 \int_\beta^{\Mf^q} \frac{\rmd s}{\lambda(s)} \le 
 \int_0^{\Mf^q} \frac{\rmd s}{\lambda(s)} <\infty,
\end{align}
since $s^{-1/q}$ is integrable at zero, due to $q>1$.  
This proves (\ref{eq:semidispsmallp}).

If $p=1$, then $p+q+1=3$, and, by Assumption \ref{th:DRass}, 
the integral over $x$ is bounded by 
$c_0 \sabs{\ln (1-a)+\ln s}^{p_0}\le c_0 2^{p_0} \sabs{\ln (1-a)}^{p_0}
\sabs{\ln s}^{p_0}$. Then the integral over $s$ 
can be estimated by 
\begin{align}
& \int_\beta^{\Mf} \frac{\rmd s}{\lambda(s)} \sabs{\ln s}^{p_0} \le 
 2 \sabs{\Mf-s_c} \max(\sabs{\ln \Mf}^{p_0},\sabs{\ln s_c}^{p_0}) 
\nonumber \\ & \qquad
 + 2 s_c \sabs{s_c-1} \sabs{\ln s_c}^{p_0} +
 \int_\beta^{1} \!\rmd s\,\frac{2 s_c}{s} \sabs{\ln s}^{p_0} 
\end{align}
where $s_c=9 \norm{\omega}'_2/2$.
By Lemma \ref{th:lnsint}, the final integral can be bounded by a constant
times $\sabs{\ln \beta}^{p_0+1}$.
Collecting the powers of $\sabs{\ln\beta}$ together, and denoting the
remaining factor by $C_0$ proves (\ref{eq:semidisp}).

\section{Suppression of crossings (Proof of  ``if'' in Theorem \ref{th:thmain})}
\label{sec:crossing}

\subsection{Uniform minimal curvature}
\label{sec:unif}

\begin{theorem}\label{th:thuniform}
Let $d\ge 2$, 
and let $\omega:\R^d\to \R$ be real-analytic and $\Z^d$-periodic. 
Then one and only one of the following alternatives is true:
\begin{enumerate}
\item\label{it:alt1} There is an affine hyperplane $M\subset \R^d$ such that 
$\omega$ is constant on $M$.
\item\label{it:alt2}  There are an integer $n_0\ge 2$ and a constant
$\vep_0>0$ with the following property:
for any $k\in \R^d$ and $u\in S^{d-1}$, there is an integer $n$
with $2\le n \le n_0$, and a direction $v\in S^{d-1}$ orthogonal to $u$, such that 
\begin{align}\label{eq:bendin}
 \frac{1}{n!} \left|(v \cdot \nabla)^n \omega(k)\right| > \vep_0 .
\end{align}
\end{enumerate}
\end{theorem}

We will use the remainder of the subsection for the proof.  From now on,
assume that $d$ and $\omega$ satisfy the assumptions of the Theorem.
Let $X=C^{\infty}(\R^d,\C)$ denote 
the topological vector space of smooth functions
equipped with its usual \frechet\ topology.
The topology is uniquely determined by the local base
given  by the sets
\begin{align}
  B^{(N)} = \defset{f\in X}{p_N(f)<\frac{1}{N}} ,
\end{align}
with $N\in \N_+$ and $p_N$ denoting the seminorm
\begin{align}
 p_N(f) = \max \defset{|D^\alpha f(x)|}{|\alpha|\le N,|x|\le N}.
\end{align} 
For more details, see \cite{rudin:fa}, section 1.46.

We recall that if $X$ and $Y$ are two topological vector spaces with 
local bases $\mathcal{B}_X$ and $\mathcal{B}_Y$, respectively, then a
function $F:X\to Y$ is continuous if and only if it has the following
property: For all $B\in \mathcal{B}_Y$ and $x\in X$ there is 
$B'\in \mathcal{B}_Y$ such that $F(x+B') \subset F(x)+B$.
From this, it is straightforward to prove the continuity of
the following two basic mappings:
for any $v\in \R^d$, the mapping $f\mapsto v\cdot \nabla f$
is a continuous, linear map $X\to X$, and
for any $x\in \R^d$ the functional $f\mapsto f(x)$ is continuous on $X$. 
Therefore, also the functional $f\mapsto (v\cdot  \nabla)^n f(0)$
is always continuous on $X$, which proves the following result.
\begin{lemma}\label{th:Uopen}
For any $n\in \N_0$, $v\in \R^d$  and $\vep\ge 0$, let
\begin{align}\label{eq:defUnv}
U_{n,v,\vep} = \defset{f\in X}{\frac{1}{n!}|(v\cdot  \nabla)^n f(0)|>\vep}.
\end{align}
Then every such $U_{n,v,\vep}$ is open in $X$.
\end{lemma}

The proof of Theorem \ref{th:thuniform} will rely on compactness of 
$S^{d-1}\times \T^d$ and on the continuity of the following auxiliary mapping.
\begin{definition}
Let $F: S^{d-1}\times \T^d \to X$ be defined, for any $x_0\in\R^d$, by
\begin{align}
F(u,[x_0])(x) = \omega(x-(x\cdot u) u + x_0)= \omega(Q_u x + x_0).
\end{align}
\end{definition}
Since $\omega$ is periodic,
$F(u,[x_0])$ does not depend on the choice of $x_0$, and, 
by smoothness of $\omega$, $F(u,[x_0])$ is also always smooth.  
Thus $F$ is a well-defined function $S^{d-1}\times \T^d \to X$, as claimed
above.  In addition, $F(u,k)$ is always 
real-analytic and constant in the
direction $u$: $F(u,k)(x+ s u)=F(u,k)(x)$ for all $s\in \R$.  Moreover,
\begin{proposition}\label{th:Fiscont}
$F$ is continuous.
\end{proposition}
\begin{proof}
Let us first show that to prove the continuity of $F$, it is enough to show
that for all
$u_0\in S^{d-1}$, $k_0\in \R^d$ and $N\in \N_+$ there is 
$\delta >0$ such that
\begin{align}
p_N(F(u,k)-F(u_0,k_0))<\frac{1}{N}
\end{align}
for all $u\in S^{d-1}$ and $k\in \R^d$ with
$|k-k_0|<\delta$ and $|u-u_0|<\delta$.  Namely, assume that 
the above condition is satisfied.  Let $V\subset X$ be open and denote 
$V_0 = F^{\gets}V$. If $V_0$ is empty, it
is trivially open.  Otherwise, let $(u_0,k_0)\in V_0$ be arbitrary, when
there is $N\in \N_+$ such that $F(u_0,k_0)+B^{(N)}\subset V$.  We
choose $\delta$ as above.  Then
$U=\defset{(u,[k])}{|u-u_0|<\delta,|k-k_0|<\delta}$ is open in 
$S^{d-1}\times \T^d$ and $F(U)\subset F(u_0,k_0)+B^{(N)}\subset V$.
Thus $V_0$ is also open, and $F$ has been proven continuous.

To prove the above property, 
we first note that for any multi-index $\alpha$ there is a finite
collection of constants $c_{\beta,\gamma}(\alpha)$, such that for all $x,u,k$
\begin{align}
D^\alpha F(u,k)(x) = \sum_{\beta:|\beta|=|\alpha|}
 \sum_{\gamma:|\gamma|\le 2|\alpha|} c_{\beta,\gamma}(\alpha) u^\gamma 
 D^\beta\omega(x-(x\cdot u) u + k);
\end{align}
this can be proven by straightforward induction in $|\alpha|$.
Therefore,
\begin{align}\label{eq:DaFdiff}
& |D^\alpha F(u,k)(x)-D^\alpha F(u_0,k_0)(x)| \le 
\sum_{\beta,\gamma} |c_{\beta,\gamma}(\alpha)| |u^\gamma-u_0^\gamma|
\norm{\omega}'_{|\alpha|}
 \\ &\qquad 
+ \sum_{\beta,\gamma} |c_{\beta,\gamma}(\alpha)| 
| D^\beta\omega(x-(x\cdot u) u + k) -
D^\beta\omega(x-(x\cdot u_0) u_0 + k_0) | . \nonumber
\end{align}

Let $\delta>0$, and choose any
$|k-k_0|<\delta$, $|u-u_0|<\delta$.  Then
by the Leibniz rule and $|u|,|u_0|=1$, we find
\begin{align}
|u^\gamma-u_0^\gamma|\le 2^{|\gamma|} \delta .
\end{align}
By expressing the difference as an integral over a derivative in the
direction of the line connecting the points, we find the estimate
\begin{align}
& | D^\beta\omega(x-(x\cdot u) u + k) -
D^\beta\omega(x-(x\cdot u_0) u_0 + k_0) | \nonumber \\ &\quad 
\le \norm{\omega}'_{|\beta|+1} 
|(x\cdot u) u-(x\cdot u_0) u_0+k-k_0| 
\end{align}
where, for all $|x|\le N$,
\begin{align}
|(x\cdot u) u-(x\cdot u_0) u_0+k-k_0| \le
2 |x| |u-u_0|+|k-k_0| \le (2 N +1)\delta.
\end{align}
By (\ref{eq:DaFdiff}) then
\begin{align}
& p_N(F(u,k)-F(u_0,k_0)) \le \delta 
( 4^{N}\norm{\omega}'_N + (2 N +1) \norm{\omega}'_{N+1} )
\max_{|\alpha|\le N} \sum_{\beta,\gamma} |c_{\beta,\gamma}(\alpha)| .
\end{align}
Since $\omega$ is periodic, $\norm{\omega}'_n <\infty$ for all $n\in \N$,
which implies that the factor multiplying $\delta$ on the right hand side is
always finite. Thus by choosing a small enough $\delta$, the bound can be made
less than $1/N$.
\end{proof}

\begin{lemma}\label{th:altern1}
Let $u\in S^{d-1}$, $k\in \R^d$ be given, and denote $f=F(u,k)$.
Then either $f$ is constant, or there is $n\ge 2$, $v\in S^{d-1}$, 
and $\vep>0$, such that $|(v\cdot  \nabla)^n f(0)|>n! \vep$.
\end{lemma}
\begin{proof}
Suppose $f$ is not constant.  Then there is $x_0\ne 0$ such that 
$f(x_0)\ne f(0)$.  
Let us define $v=x_0/|x_0|$, when $v\in S^{d-1}$,
and let $g:\R\to \R$ be defined by 
$g(t)=f(t v)-t v\cdot\nabla f(0)-f(0)$.  Then $g$ is real-analytic with
$g(0)=0$ and $g'(0)=0$.  If $g^{(n)}(0)= 0$ for all $n\ge 2$, then 
$g=0$ everywhere, i.e., $f(t v)=t v\cdot\nabla f(0)+f(0)$ for all $t\in \R$.
Since $f(|x_0| v)\ne f(0)$, then necessarily $v\cdot\nabla f(0)\ne 0$, and 
thus $\lim_{t\to \infty} |f(t v)| = \infty$.  However, this contradicts the
obvious bound $\norm{f}_0 \le \norm{\omega}'_0 < \infty$, and thus
we can conclude that there is $n\ge 2$ such that 
$g^{(n)}(0)=(v\cdot  \nabla)^n f(0) \ne 0$.  Thus, for instance,
$\vep=|(v\cdot  \nabla)^n f(0)|/(2 n!)>0$ suffices for the bound in the Lemma.
\end{proof}
\begin{proofof}{Theorem \ref{th:thuniform}}
Let us first note that for any $u\in S^{d-1}$, $x_0\in \R^d$,
the image of $x\mapsto x-(x\cdot u) u + x_0$ is exactly the
affine hyperplane $\defset{x\in \smash{\R^d}}{x\cdot u = x_0\cdot u}$.
Thus the first alternative is true if and only if there is
$u\in S^{d-1}$, $x_0\in \R^d$ such that
$F(u,k)$, $k=[x_0]$, is constant.  On the other hand, then
also $g_v(t)=F(u,k)(t v) = \omega(tQ_u v + x_0)$ is constant
for any $v\in S^{d-1}$, and 
thus $0=g_v^{(n)}(0) = 
(Q_{u} v \cdot \nabla)^n \omega(k)$ for all $n\ge 2$.
This proves that
the second alternative is false when the first is true.

Denote $K=S^{d-1}\times \T^d$ which is compact when endowed 
with the metric $d((u,k),(u',k'))=|u-u'|+\torusd(k,k')$.
Suppose that the first alternative is false, when we know that 
$F(u,k)$ is never a constant.  By Lemma \ref{th:altern1}, then
\begin{align}
F(K) \subset \bigcup_{n,v,\vep} U_{n,v,\vep}
\end{align}
where $U_{n,v,\vep}$ is defined by (\ref{eq:defUnv}), and
the union is taken over all 
$n\in \N$, with $n\ge 2$, and $v\in S^{d-1}$, $\vep>0$.  
Since $K$ is compact and, by 
Proposition \ref{th:Fiscont}, $F$ is continuous, 
$F(K)$ is compact.  Using Lemma \ref{th:Uopen}, we can thus conclude 
that $U_{n,v,\vep}$ form an open cover of the compact set $F(K)$.
Therefore, there
is a finite sequence $(n_i,v_i,\vep_i)$ such that $(U_{n_i,v_i,\vep_i})$
cover the whole image of $F$.  
Let
\begin{align}
 \vep_0 = \min_i \vep_i>0\qand n_0 = \max_i n_i\ge 2.
\end{align}

Let $u\in S^{d-1}$, $x_0\in \R^d$ be
arbitrary, and let $k=[x_0]$.  There is an index $i$ such that
$f=F(u,k) \in U_{n,v',\vep_i}$, with $v'=v_i\in S^{d-1}$ and
$n=n_i$.   Then $2\le n\le n_0$, and
we have $|(v'\cdot \nabla)^n f(0)|> n! \vep_i \ge n! \vep_0$. Since
$(v'\cdot \nabla)^n f(x) = 
\left.(Q_{u} v'\cdot \nabla)^n \omega\right|_{Q_{u}x+x_0}$,
we have also  $|(v\cdot \nabla)^n \omega(x_0)|> n! \vep_0$ with
$v = Q_{u} v'/|Q_{u} v'|$ (note that obviously $|Q_{u} v'|\ne 0$).
As $v\cdot u=0$,  
the pair $n,v$ has the properties required by the second alternative.
\end{proofof}

\subsection{Crossing estimate}

Let us assume that $\omega$ is not a constant on any affine hyperplane.
Then we can find  constants $n_0\ge 2$ and $0<\vep_0\le \frac{1}{2}$,
for which
the second alternative in Theorem \ref{th:thuniform} holds.
As in Proposition \ref{th:psiderivprop}, 
let $M_n=\norm{\omega}'_n$, $a_0=\max(1,8 M_2)$, and define
\begin{align}
 \mu = \frac{1}{1+ 2^{n_0+3} + M_{n_0+1} 2^{4 n_0+1}}
\end{align}
when $0<\mu\le \frac{1}{33}$, and $\mu$ satisfies the conditions of 
the Proposition for any $2\le N\le n_0$.
We also define for any given
$0<r\le 1$ and $2\le N\le n_0$,
\begin{align}
\vep(r,N) = \vep_0 (r \mu)^{n_0-N} \le \vep_0\le \frac{1}{2}.
\end{align}

Consider arbitrary given $k_0\in \T^d$, $\alpha\in \R^3$, and $0<\beta\le
1$.  We 
need to estimate
\begin{align}
&I=\scint(\alpha,k_0,\beta) =   \int_{(\T^d)^2} \rmd k_1\rmd k_2\, 
\prod_{j=1}^3 \frac{1}{|\alpha_j-\omega(k_j)+\ci\beta|}
\end{align}
where $k_3= k_3(k_1,k_2)= k_1-k_2+k_0$.  By using a layer cake representation,
\begin{align}
& I= \int_0^{\Mf}\! \rmd s \int_{(\T^d)^2} \rmd k_1\rmd k_2\, 
\prod_{j=1}^3 \frac{1}{|\alpha_j-\omega(k_j)+\ci\beta|}
\frac{\1(\min_j |\nabla \omega(k_j)|\ge s) }{\min_j |\nabla \omega(k_j)|}
\nonumber \\ & \quad
\le  3 \int_0^{\beta^\gamma}\! \frac{ \rmd s}{\beta}
 \int_{\T^d}\! \frac{\rmd k'}{|\alpha -\omega(k')+\ci\beta|} 
 \int_{\T^d}\! \frac{\rmd k}{|\nabla\omega(k)|}
\frac{1}{|\alpha -\omega(k)+\ci\beta|} 
 \nonumber \\ & \qquad
+  \int_{\beta^\gamma}^{\Mf}\! \rmd s\!
 \int_{(\T^d)^2} \rmd k_1\rmd k_2\, 
\prod_{j=1}^3 \frac{1}{|\alpha_j-\omega(k_j)+\ci\beta|}
\frac{\1(\min_j |\nabla \omega(k_j)|\ge s) }{\min_j |\nabla \omega(k_j)|}
\end{align}
where, to get the first term, 
we have used $\frac{1}{\min_j |\nabla \omega(k_j)|}
\le \sum_j \frac{1}{|\nabla \omega(k_j)|}$ and then estimated one of
the factors trivially, followed by a change of variables.
By Theorem \ref{th:semidisp}, the first term is bounded by
\begin{align}
3 C_0 C_1 \sabs{\ln\beta}^{p_0+3} \beta^{-(1-\gamma)}
\end{align}
and so it is ``harmless'' for any $\gamma>0$.

To estimate the second term,
let us define for any $s>0$,
\begin{align}\label{eq:deflandd}
r_0(s) = \min(1,\frac{s}{2}),\quad
\lambda(s) = \frac{1}{4}\min \Big(\frac{1}{2},
\frac{\vep_0}{a_0} (r_0(s) \mu)^{n_0}\Bigr)
\quad\text{and}\ \ \delta(s) = \lambda(s).
\end{align}
We use the cut-off function
$G$ as before inside the $k$ integrals,
\begin{align}
I_2 & =  \int_{(\T^d)^2} \rmd k_1\rmd k_2\, 
\prod_{j=1}^3 \frac{1}{|\alpha_j-\omega(k_j)+\ci\beta|}
\frac{\1(\min_j |\nabla \omega(k_j)|\ge s) }{\min_j |\nabla \omega(k_j)|}
 \nonumber \\ &
 =  \int_{(\T^d)^2} \rmd x_1\rmd x_2\, 
 \int_{(\T^d)^2} \rmd k_1\rmd k_2\, \prod_{j=1}^2
\frac{G(x_j-k_j,\lambda(s))}{|\alpha_j -\omega(k_j)+\ci\beta|} 
 \nonumber \\ & \quad \times
\frac{1}{|\alpha_3-\omega(k_3)+\ci\beta|}
\frac{\1(\min_j |\nabla \omega(k_j)|\ge s) }{\min_j |\nabla \omega(k_j)|} .
\end{align}
Let $x_3=x_1-x_2+k_0$, when  
inside the integral, for $j=1,2,3$,
\begin{align}
&  |\, |\nabla\omega(k_j)| - |\nabla\omega(x_j)|\, | \le 
  |k_j-x_j| \norm{\omega}'_2 \le 2 \lambda \norm{\omega}'_2
  \le \frac{s}{2} \le \frac{|\nabla\omega(k_j)|}{2},
\end{align}
since $|k_j-x_j| \le \lambda < 2\lambda$ for
$j=1,2$, and $|k_3-x_3| \le 2\lambda$. 
Therefore, we have
$|\nabla \omega(x_j)|\ge \frac{1}{2} |\nabla \omega(k_j)|\ge \frac{s}{2}$
and $2|\nabla \omega(k_j)|\ge |\nabla \omega(x_j)|$,
for all $j$, and thus
\begin{align}
 I_2 & \le 
 2 \int_{(\T^d)^2} \rmd x_1\rmd x_2\,
\frac{\1(\min_j |\nabla \omega(x_j)|\ge \frac{1}{2}s)}{
  \min_j |\nabla \omega(x_j)|} 
 \nonumber \\ & \quad \times
 \int_{(\T^d)^2} \rmd k_1\rmd k_2\, \prod_{j=1}^2
\frac{G(x_j-k_j,\lambda)}{|\alpha_j -\omega(k_j)+\ci\beta|} 
\frac{1}{|\alpha_3-\omega(k_3)+\ci\beta|} .
\end{align}
In particular, now inside the $x$ integrals we have
$r_0(s)\le\min(1,|\nabla\omega(x_j)|)$,  for all $j=1,2,3$,
and, since $\lambda(s)\le r_0(s)/a_0$, we can apply the
results of Proposition \ref{th:psiderivprop} around any of the points $x_j$.

We next need to estimate, for given $\alpha\in \R^3$ and $x_j\in \T^d$
the integral
\begin{align}\label{eq:defJint}
 J = 
 \int_{(\T^d)^2} \rmd k_1\rmd k_2\, \prod_{j=1}^2
\frac{G(x_j-k_j,\lambda)}{|\alpha_j -\omega(k_j)+\ci\beta|} 
\frac{1}{|\alpha_3-\omega(k_3)+\ci\beta|} ,
\end{align}
assuming $\min_j |\nabla \omega(x_j)|\ge s/2\ge \beta^{\gamma}/2>0$.
Since then $\nabla\omega(x_j)\ne 0$, we can define
for all $j=1,2,3$, 
\begin{align}
 u_j =\frac{\nabla\omega(x_j)}{|\nabla\omega(x_j)|} \in S^{d-1}.
\end{align}
We apply different estimates depending on whether all $u_j$ are almost
parallel or not.  A sufficient degree of separation turns out to be 
determined by the 
parameter $\delta$ defined in (\ref{eq:deflandd}), for which in 
particular $0<\delta\le 
\frac{1}{2}$. 
The first of the estimates is applied, if
\begin{align}\label{eq:nonparall}
|u_1\cdot u_{3}|\le \sqrt{1-\delta^2} \qquad\text{or}\qquad
|u_2\cdot u_{3}|\le \sqrt{1-\delta^2},
\end{align}
and otherwise the second estimate is used.  

In subsection \ref{sec:nonpar} we shall prove
that in the first case there is a constant 
$C'_1$, depending only on $\omega$, such that
\begin{align}\label{eq:JCp1bound}
 J & \le \frac{\sabs{\ln\beta}^3}{\delta(s) \lambda(s)^3}
\frac{C_1'}{\prod_{j=1}^3 |\nabla\omega(x_j)|}.
\end{align}
The other, more involved estimate, is done in subsection 
\ref{sec:parallel}.  There we prove that, if we choose 
\begin{align}\label{eq:defgammafin}
  \gamma = \frac{1}{3 n_0 (n_0+1)},
\end{align}
then in the second case
there are constants $C_2'$ and $\beta_0$, depending only on $\omega$,
such that for all $0<\beta\le \beta_0$,
\begin{align}\label{eq:JCp2bound}
 J & \le \frac{\sabs{\ln\beta}^2}{\lambda(s)^3}  \beta^{\frac{1}{n_0+1}-1}
\frac{C_2'}{\prod_{j=1}^3 |\nabla\omega(x_j)|}.
\end{align}
After applying either one of the inequalities, the remaining
integral over $x_j$ can be estimated using the Young inequality,
\begin{align}
& \int_{(\T^d)^2} \rmd x_1\rmd x_2\,
\frac{\1(\min_j |\nabla \omega(x_j)|\ge \frac{1}{2}s)}{
  \min_j |\nabla \omega(x_j)|} 
\frac{1}{\prod_{j=1}^3 |\nabla\omega(x_j)|}
 \nonumber \\ & \quad
\le \sum_{j=1}^3 \int_{(\T^d)^2} \rmd x_1\rmd x_2\,
\frac{\1(|\nabla \omega(x_j)|\ge \frac{1}{2}s)}{|\nabla \omega(x_j)|^2} 
\prod_{j'=1;j'\ne j}^3 \frac{1}{|\nabla\omega(x_{j'})|}
 \nonumber \\ & \quad
\le 3 \Bigl(  \int_{\T^d} \rmd x\,
\frac{\1(|\nabla \omega(x)|\ge \frac{1}{2}s)}{|\nabla \omega(x)|^3}
\Bigr)^{2/3} \Bigl( \int_{\T^d} \rmd x\,
\frac{1}{|\nabla \omega(x)|^{3/2}}\Bigr)^{4/3} .
\end{align} 
Thus by Lemma \ref{th:L1lemma} and Assumption  \ref{th:DRass}, 
there is a constant $C'$ such that for all sufficiently small $\beta$
\begin{align}\label{eq:I2sbound}
&  I_2(s) \le C' \sabs{\ln\beta}^{\frac{2}{3}p_0 + 3}\left(
 \lambda(s)^{-4}+ \lambda(s)^{-3}  \beta^{\frac{1}{n_0+1}-1} 
\right) .
\end{align}

If $s\ge 2$, then $r_0=1$ and $\lambda(s)$ is equal to a non-zero constant,
implying that the bound in (\ref{eq:I2sbound}) is independent of $s$, and
the integral over $2\le s\le \Mf$ is thus easily estimated.
If $0<s\le 2$, we have $r_0=s/2$, and thus for these $s$,
\begin{align}
\lambda(s) = \frac{\vep_0}{4 a_0} (\mu s /2)^{n_0}
 \ge 2^{-2-n_0} \vep_0 \mu^{n_0} s^{n_0}.
\end{align}
Therefore, there is $C''$ such that
\begin{align}
 \int_{\beta^\gamma}^{2}\! \rmd s\,
 I_2(s) \le C'' \sabs{\ln\beta}^{\frac{2}{3}p_0 + 3}\left(
 \beta^{-\gamma (4 n_0-1)} +
 \beta^{-\gamma (3 n_0-1)+\frac{1}{n_0+1}-1} 
\right) ,
\end{align}
where, by our choice of $\gamma$,
\begin{align}
& 1-\gamma (4 n_0-1) = \gamma (3n_0 ^2-n_0+1) \ge \gamma
\ \text{ and } \
-\gamma (3 n_0-1)+\frac{1}{n_0+1} =  \gamma.
\end{align}
Collecting all the results together, we  
have now proven that there are constants $\beta_0$ and $C$, depending only
on $\omega$, 
such that for all $0<\beta\le \beta_0$, $\alpha\in \R$,  and $k_0\in \T^{d}$,
\begin{align}
  \scint(\alpha,k_0,\beta) \le C
 \sabs{\ln\beta}^{p_0 + 3} \beta^{\gamma-1} .
\end{align}
For $\beta \ge \beta_0$, we can trivially estimate 
$\scint(\alpha,k_0,\beta) \le \beta_0^{-3}$, which allows us to conclude that
the dispersion relation suppresses crossings with a power  of (at least)
$\gamma$.  Therefore, we only need to derive the estimates
(\ref{eq:JCp1bound}) and 
(\ref{eq:JCp2bound}) to complete the proof of the Theorem \ref{th:thmain}.

We proved the result for $\gamma$ defined in (\ref{eq:defgammafin}).  This
value is not 
optimal, as shown by the one example for which the power has previously been
estimated, that is, 
for the nearest neighbor interaction in $d=3$.  The
corresponding dispersion relation is a Morse function, but there are points
at which its Hessian vanishes.  Thus we need to take at least $n_0=3$
above, which would yield $\gamma \le \frac{1}{36}$.  However, 
in \cite{erdyau04} it has been proven that a power
$\gamma=\frac{1}{4}$ can be allowed for this case.

\subsubsection{Non-parallel gradients}
\label{sec:nonpar}

We assume in this subsection that (\ref{eq:nonparall}) holds,
meaning that $u_3$ is not nearly parallel to
one of the vectors $u_1$ or $u_2$, say to the vector $u_1$.
This will allow us to estimate the
$k_3$ factor in the crossing integral by integrating it out in the
direction determined by the projection of $u_3$ orthogonal to the level set 
of $\omega$ at $k_1$.  As we will show next in more detail,
this will prove the estimate given in
(\ref{eq:JCp1bound}).

Consider the integral defining $J$ in (\ref{eq:defJint}).
We first change the integration variables in the following manner:
if $|u_1\cdot u_{3}|\le \sqrt{1-\delta^2}$, we 
define $k'_1 = k_1-x_1$ and $k'_2 = k_2-x_2$, when 
$k_3 = k'_1-k'_2+x_3$. Otherwise, 
$|u_2\cdot u_{3}|\le \sqrt{1-\delta^2}$, and we define
$k'_1 = k_2-x_2$ and $k'_2 = k_1-x_1$, when
$k_3 = -(k'_1-k'_2)+x_3$.  
It is thus enough derive the bound assuming
$|u_1\cdot u_{3}|\le \sqrt{1-\delta^2}$, if we allow slightly more general
dependence of $k_3$ on the integration variables, namely if we assume
$k_3= x_3 + \sigma(k'_1-k'_2)$, with $\sigma=\pm 1$ (swapping the indices
$1\leftrightarrow 2$ in the result then produces the corresponding bound
for the second case).  

As mentioned already before, 
$\lambda$ is small enough that we can apply
Lemma \ref{th:standifth} and obtain two diffeomorphisms
$\psi_1$ and $\psi_2$
such that Corollary \ref{th:psicoroll} holds.  This shows that
\begin{align}
 J & \le 2^2 \int_{|y|<2\lambda} \!\rmd y
 \int_{|y'|<2\lambda} \!\rmd y'\, \frac{N_d^2}{\lambda^{2d}}
\frac{1}{|\alpha_1 -\omega(x_1)-|\nabla\omega(x_1)| y_1+\ci\beta|} 
\nonumber \\ & \qquad\times 
\frac{1}{|\alpha_2 -\omega(x_2)-|\nabla\omega(x_2)| y'_1+\ci\beta|} 
\frac{1}{|\alpha_3-\omega(x_3+\gamma(y',y))+\ci\beta|}
\end{align}
where
\begin{align}
\gamma(y',y) = \sigma( \psi_1(y)-x_1 - (\psi_2(y')-x_2) ).
\end{align}
By Lemma \ref{th:standifth}, always
\begin{align}\label{eq:gammabound}
  |\gamma(y',y)| \le |\psi_1(y)-x_1| +|\psi_2(y')-x_2| <  8\lambda.
\end{align}

Let us denote $A=D\psi_1(0)$, which is a rotation in $\R^d$
with $A^T u_1 = e_1$.
Let $v=Q_{u_1} u_3$ and $v'= A^T v$.  By assumption then
\begin{align}
 |v|^2 = 1- (u_1\cdot u_{3})^2 > \delta^2
\end{align}
implying that $|v'|= |v| > \delta>0$.  Also, $v'_1= 0$ since 
$v\cdot u_1 =0$.  
Thus there is a rotation $O$ of $\R^d$ for which
$Oe_1 = e_1$ and $Ov' = |v'| e_2$.  We change the integration variable
$y$ to $z= O y$, yielding
\begin{align}\label{eq:Jest1}
 J & \le \frac{2^2 N_d^2}{\lambda^{2d}} \int_{|z|<2\lambda} \!\rmd z
 \int_{|y'|<2\lambda} \!\rmd y'\, 
\frac{1}{|\alpha_1 -\omega(x_1)-|\nabla\omega(x_1)| z_1+\ci\beta|} 
\nonumber \\ & \quad\times 
\frac{1}{|\alpha_2 -\omega(x_2)-|\nabla\omega(x_2)| y'_1+\ci\beta|} 
\frac{1}{|\alpha_3-\omega(x_3+\gamma(y',O^T z))+\ci\beta|}
\nonumber \\ &
 \le \frac{2^2 N_d^2}{\lambda^{2d}} \frac{(2 \lambda)^{2 d-3}}{N_{d-2}N_{d-1}}
\int_{|z_1|<2\lambda} \!\rmd z_1\,
\frac{1}{|\alpha_1 -\omega(x_1)-|\nabla\omega(x_1)| z_1+\ci\beta|} 
\nonumber \\ & \quad\times 
 \int_{|y'_1|<2\lambda} \!\rmd y'_1\, 
\frac{1}{|\alpha_2 -\omega(x_2)-|\nabla\omega(x_2)| y'_1+\ci\beta|} 
\nonumber \\ & \quad\times 
\sup_{\substack{|z|,|y'|< 2\lambda \\ z_2 =0}}
\, \int_{t^2<(2\lambda)^2-z^2} \!\rmd t\,
\frac{1}{|\alpha_3-\omega(x_3+\gamma(y',O^T (z+t e_2)))+\ci\beta|} .
\end{align}

Let us first estimate the final term, of the form
\begin{align}
\int_{|t|<R} \!\rmd t\, \frac{1}{|\alpha_3-f(t)+\ci\beta|},
\end{align}
where
\begin{align}
f(t) = \omega(x_3+ \Gamma(t) ), \quad \text{with}\quad
\Gamma(t) = \gamma(y',O^T\! z+ t O^T\! e_2) .
\end{align}
Clearly, $|f|\le \norm{\omega}'_0<\infty$, and we shall later 
show that
\begin{align}\label{eq:fprimeest}
& | f'(t) | \ge  
\frac{1}{4} |\nabla \omega (x_3)| \delta > 0.
\end{align}
This allows applying Lemma \ref{th:genfboundn1}, and proves that
\begin{align}
\int_{|t|<R} \!\rmd t\, \frac{1}{|\alpha_3-f(t)+\ci\beta|}
\le \frac{4\cdot 6 \sabs{\ln \sabs{\norm{\omega}'_0}}}{
  |\nabla \omega (x_3)| \delta} \sabs{\ln \beta}.
\end{align}
Therefore, applying Proposition \ref{th:polybound} to (\ref{eq:Jest1}),
yields the bound\begin{align}
 J & \le \frac{2^{2 d+1} N_d^2}{N_{d-2}N_{d-1}}
\frac{6^3\sabs{\ln\beta}^3}{\delta \lambda^3}
\frac{\sabs{\ln \sabs{\norm{\omega}'_0}}
\sabs{\ln \sabs{\norm{\omega}'_1}}^2
}{\prod_{j=1}^3 |\nabla\omega(x_j)|}.
\end{align}
Collecting all the constants into $C'_1$ proves that 
(\ref{eq:JCp1bound}) is valid in this case.

We still need to prove (\ref{eq:fprimeest}).  From the definition of $f$,
\begin{align}
  f'(t) = \Gamma'(t)\cdot \nabla \omega (x_3+\Gamma(t)).
\end{align}
Since $O^T e_2 = v'/|v'| = A^T v /|v|$,
\begin{align}
\Gamma(t) = \sigma (\psi_1(O^T\! z+ t A^T v/|v|)-x_1) 
- \sigma (\psi_2(y')-x_2) ,
\end{align}
which, together with (\ref{eq:Dpsieq}), implies
\begin{align}
  \Gamma'(t) = \frac{\sigma}{|v|} 
  \left.  D \psi_1\right|_{O^T\! z+ t v'/|v|} A^T v 
 = \frac{\sigma}{|v|} \Bigl( v - 
\frac{\nabla\omega(x)\cdot v}{\nabla\omega(x)\cdot u_1}
 u_1 \Bigr) 
\end{align}
where $x=x(t)=\psi(O^T\! z+ t A^T v/|v|)$, and 
$ |\nabla \omega(x)-\nabla \omega(x_1)| \le 
 \frac{1}{2} |\nabla \omega(x_1)|$. 
Therefore,
\begin{align}
&  f'(t)  = \frac{\sigma}{|v|} \Bigl( v\cdot\nabla \omega (x_3) +
v\cdot[\nabla \omega (x_3+\Gamma(t)) -\nabla \omega (x_3) ] 
\nonumber \\ & \qquad
-  \frac{\nabla\omega(x)\cdot v}{\nabla\omega(x)\cdot u_1}
 u_1 \cdot \nabla \omega (x_3+\Gamma(t))  \Bigr).
\end{align}
Here $v\cdot\nabla \omega (x_3) = |\nabla \omega (x_3)| v \cdot u_3$,
and $v \cdot u_3= 1-(u_1 \cdot u_3)^2 = v^2$.
Thus
\begin{align}
& | f'(t) | \ge  |\nabla \omega (x_3)| |v| - 
\Bigl( |\nabla \omega (x_3+\Gamma(t)) -\nabla \omega (x_3)| 
\nonumber \\ & \qquad
+ \frac{|\nabla\omega(x)\cdot v|}{|v||\nabla\omega(x)\cdot u_1|}
 |u_1 \cdot \nabla \omega (x_3+\Gamma(t)) | \Bigr).
\end{align}

By (\ref{eq:gammabound}), $|\Gamma(t)| <  8\lambda$, and thus
\begin{align}
& |\nabla \omega (x_3+\Gamma(t)) -\nabla \omega (x_3)| 
\le \norm{\omega}'_2 |\Gamma(t)| < 8 \norm{\omega}'_2 \lambda = \lambda'.
\end{align}
In addition,
\begin{align}
 u_1 \cdot \nabla \omega (x_3+\Gamma(t))
 = |\nabla \omega (x_3)| u_1 \cdot u_3 +
 u_1\cdot[ \nabla \omega (x_3+\Gamma(t)) -\nabla \omega (x_3)],
\end{align}
yielding $| u_1 \cdot \nabla \omega (x_3+\Gamma(t))| \le
|\nabla \omega (x_3)| +  \lambda'$.
As $u_1\cdot v=0$, now
$\nabla\omega(x)\cdot v = [\nabla\omega(x)-\nabla\omega(x_1)]\cdot v$, 
and so
\begin{align}
|\nabla\omega(x)\cdot v| \le |v| |x-x_1| \norm{\omega}'_2
< |v| \frac{1}{2}\lambda'.
\end{align}
Similarly, 
$|\nabla\omega(x)\cdot u_1| \ge |\nabla\omega(x_1)| - 
 \frac{1}{2}\lambda'$.
Since $\lambda' = 8 \norm{\omega}'_2 \lambda \le
 \frac{\delta s}{8} \le \frac{\delta}{4}  |\nabla\omega(x_j)|$,
we can conclude that
\begin{align}
\frac{|\nabla\omega(x)\cdot v|}{|v||\nabla\omega(x)\cdot u_1|}
\le \frac{\lambda'}{|\nabla\omega(x_1)|} \le \frac{\delta}{4}.
\end{align}
Therefore,
\begin{align}
& | f'(t) | \ge  |\nabla \omega (x_3)| |v| - 
\frac{\delta}{4}  |\nabla\omega(x_3)|
- \frac{\delta}{4} |\nabla \omega (x_3)| \Bigl(1 +  \frac{\delta}{4}\Bigr) 
\nonumber \\ & \quad
\ge |\nabla \omega (x_3)| (|v|- \frac{3}{4} \delta)
\ge |\nabla \omega (x_3)| \frac{1}{4} \delta > 0,
\end{align}
and we have arrived at the estimate (\ref{eq:fprimeest}).

\subsubsection{Nearly parallel gradients}
\label{sec:parallel}

We assume here 
that (\ref{eq:nonparall}) is not true, i.e., that all three of the vectors
$u_j$ are nearly parallel to each other.  In this case, we cannot integrate
the $k_3$ term in the direction of its gradient.  Instead, we will show
here that we can find a direction essentially orthogonal to the gradient, in
which the $k_3$ resolvent can be integrated, and will lead to some degree
of extra decay.  The extra decay will be caused by the higher order
curvature of the level sets.  However, we need to choose the point $k_j$ and
the direction of integration carefully, in order to make sure that the known
curvature is the dominant effect.  In particular, we cannot any more
consider the two $d$-dimensional integrals independently, but have to
choose the direction in the full $2d$-dimensional space.
Since we need to consider higher order curvature effects, we will need here
the full machinery of the technical Lemmas.

By assumption, 
$|u_j\cdot u_{3}|>\sqrt{1-\delta^2}$ for both $j=1,2$. 
For any $u,v\in S^{d-1}$, by direct computation
\begin{align}
  |Q_u v| = \sqrt{1-(u\cdot c)^2} .
\end{align}
Since
\begin{align}
u_1\cdot u_{2} =  ( (u_1\cdot u_{3}) u_3 +Q_{u_3} u_1 ) \cdot u_{2} 
= (u_1\cdot u_{3}) (u_2\cdot u_{3}) + ( Q_{u_3} u_1 ) \cdot ( Q_{u_3} u_2 ),
\end{align}
and $|Q_{u_3} u_j| = \sqrt{1- (u_3\cdot u_j)^2} < \delta$,
we have
\begin{align}
 |u_1\cdot u_{2}| \ge  |u_1\cdot u_{3}| |u_2\cdot u_{3}|
  - |Q_{u_3} u_1| |Q_{u_3} u_2| > 1- 2 \delta^2.
\end{align}
Using the fact that $\sqrt{1-x}\ge 1- 2 x$ for 
all $0\le x \le \frac{1}{4}$ to cover the other cases, we can conclude that
for all $j,j'\in\set{1,2,3}$
\begin{align}\label{eq:parall}
|u_j\cdot u_{j'}|>1- 2 \delta^2 \ge \frac{1}{2}.
\end{align}

Since we assume here that
$\omega$ is not a constant on any affine hyperplane, the second alternative
of Theorem \ref{th:thuniform} is valid, and we can thus find 
$v_3\in S^{d-1}$ such that $v_3\cdot u_3=0$ and
for some $2\le \bar{n}\le n_0$,
\begin{align}
 \frac{1}{\bar{n}!}
 \left|(v_3 \cdot \nabla)^{\bar{n}} \omega(x_3)\right| > \vep_0
\ge \vep(r_0(s),\bar{n}).
\end{align}
Let $\bar{v}_j = Q_{u_j} v_3$ for $j=1,2$. Since
\begin{align}
 |u_j \cdot v_3| = |Q_{u_3} u_j \cdot v_3|\le |Q_{u_3} u_j|
 = \sqrt{1-(u_j\cdot u_3)^2} <\delta,
\end{align}
then $|\bar{v}_j|=\sqrt{1-(u_j\cdot v_3)^2}>\sqrt{1-\delta^2}>0$.  
Therefore, we can define further
$v_j = \bar{v}_j/|\bar{v}_j|\in S^{d-1}$, when 
$v_j\cdot v_3= |\bar{v}_j|>\sqrt{1-\delta^2}$.  This 
implies, by the same argument as for $u_j$, that 
for all $j,j'\in\set{1,2,3}$,
$v_j\cdot v_{j'}>1- 2 \delta^2$, and thus also
\begin{align}\label{eq:vjjdist}
|v_j-v_{j'}| = \sqrt{2(1-v_j\cdot v_{j'})}<2\delta.
\end{align}

We have now constructed unit
vectors $v_j$, $j=1,2,3$, such that $u_j\cdot v_j=0$.
For each $j$ let us associate an integer $n_j$ which is the 
smallest of integers $n\ge 2$ for which 
\begin{align}
 \frac{1}{n!}
\left|(v_j \cdot \nabla)^{n} \omega(x_j)\right| > \vep(r_0(s),n);
\end{align}
if no such integer exists, let $n_j=\infty$.
Let $j_0$ be an index which has the smallest $n_j$, 
and denote $N=n_{j_0}$.  Since
$n_3 \le\bar{n}\le n_0$, we know that $2\le N\le n_0<\infty$.
Let $\vep = 2 \vep(r_0,N)= 2 \vep_0 (r_0 \mu)^{n_0-N}\le 1$. Then, 
for any $2\le n<N$ and $j=1,2,3$, by construction we have
\begin{align}\label{eq:firstomest}
 \frac{1}{n!}
 \left|(v_j \cdot \nabla)^{n} \omega(x_j)\right| \le \vep(r_0,n)
 = \frac{1}{2}\vep (r_0 \mu)^{N-n}  ,
\end{align}
and $ \frac{1}{N!}\left|(v_{j_0} \cdot \nabla)^{N} \omega(x_{j_0})\right| 
 > \frac{1}{2}\vep$.

Let $\pi$ be the unique 
cyclic permutation of the indices $(1,2,3)$ for which $\pi(3)=j_0$,
and let us define $k'_j = k_{\pi(j)}$, and permute $\alpha$,
$x$, $u$ and $n$ similarly to yield $\alpha'$,
$x'$, $u'$ and $n'$.
We change the integration variables from $(k_1,k_2)$ to $(k'_1,k'_2)$.
This modifies the functional dependence of $k'_3$ on the integration
variables: for $j_0=3$, $k'_3= k'_1-k'_2+k_0$, 
for $j_0=2$, $k'_3= k'_2-k'_1+k_0$, 
for $j_0=1$, $k'_3= k'_1+k'_2-k_0$.  (The dependence of $x'_3$ on
$x'_1$, $x'_2$, and $k_0$, changes accordingly with $j_0$.)
On the other hand, since 
$|k_j-x_j|< \lambda$, the new integration region is contained in 
$|k'_1-x'_1|,|k'_2-x'_2|< 2\lambda$. Thus if we can derive a bound for
$J'=\sup_{\sigma\in \set{\pm 1}^3} J'(\sigma)$,
\begin{align}
 J'(\sigma) = \frac{N_d^2}{\lambda^{2 d}}
 \int_{(\T^d)^2} \rmd k'_1\rmd k'_2\, \prod_{j=1}^2
\frac{ \1(|k'_j-x'_j|<2 \lambda)}{|\alpha'_j -\omega(k'_j)+\ci\beta|} 
\frac{1}{|\alpha'_3-\omega(k'_3)+\ci\beta|},
\end{align}
assuming $j_0=3$, $k_3=\sigma_1 k_1 + \sigma_2 k_2+ \sigma_3 k_0$, and
$x_3=\sigma_1 x_1 + \sigma_2 x_2+ \sigma_3 k_0$,
a bound for $J$ can be obtained by undoing the permutation of the
indices appropriately.

Since $2\lambda\le |\nabla\omega(x_j)|/a_0$, for all $j=1,2,3$,
we can apply Lemma \ref{th:standifth} and Corollary \ref{th:psicoroll}
to both of the $k$-integrals.  We denote the corresponding diffeomorphisms
by $\psi_1$ and $\psi_2$, and obtain the bound
\begin{align}
& J'(\sigma)  \le \frac{2^2 N_d^2}{\lambda^{2 d}}
\int_{|y|<4\lambda} \!\rmd y
\int_{|y'|<4\lambda} \!\rmd y'\,
\frac{1}{|\alpha_1 -\omega(x_1)-|\nabla\omega(x_1)| y_1+\ci\beta|} 
\nonumber \\ & \qquad\times 
\frac{1}{|\alpha_2 -\omega(x_2)-|\nabla\omega(x_2)| y'_1+\ci\beta|} 
\frac{1}{|\alpha_3-\omega(x_3+\gamma(y',y))+\ci\beta|}
\end{align}
where
\begin{align}
\gamma(y',y) = \sigma_1 ( \psi_1(y)-x_1) + \sigma_2 (\psi_2(y')-x_2) ),
\end{align}
and by Lemma \ref{th:standifth}, always 
\begin{align}
|\psi_1(y)-x_1|, |\psi_2(y')-x_2| < 2^3\lambda \qand
|\gamma(y',y)| < 2^4\lambda .
\end{align}

For $j=1,2$, let us denote the rotation $D\psi_j(0)$ by $A_j$ and define
$\tilde{v}_j = A_j^T v_j$.  Then $\tilde{v}_j\cdot e_1 = v_j\cdot u_j =0$, and 
there is a rotation $O_j$ of $\R^d$ for which
$O_j e_1 = e_1$ and $O_j \tilde{v}_j = e_2$.  We change variables to $z=O_1 y$
and $z'=O_2 y'$, and evaluate first the $z_2$ and $z'_2$ integrals.
This yields
\begin{align}
& J'(\sigma)  \le \frac{2^2 N_d^2}{\lambda^{2 d}}
\int_{|z|<4\lambda} \!\rmd z
\int_{|z'|<4\lambda} \!\rmd z'\,
\frac{1}{|\alpha_1 -\omega(x_1)-|\nabla\omega(x_1)| z_1+\ci\beta|} 
\nonumber \\ & \qquad\times 
\frac{1}{|\alpha_2 -\omega(x_2)-|\nabla\omega(x_2)| z'_1+\ci\beta|} 
\frac{1}{|\alpha_3-\omega(x_3+\gamma(O_2^T z',O_1^T z))+\ci\beta|}
\nonumber \\ & \quad
  \le \frac{2^2 N_d^2}{\lambda^{2 d}} \frac{(4\lambda)^{2(d-2)}}{N_{d-2}^2}
\int_{|z_1|<4\lambda} \!\rmd z_1\,
\frac{1}{|\alpha_1 -\omega(x_1)-|\nabla\omega(x_1)| z_1+\ci\beta|} 
\nonumber \\ & \qquad\times 
\int_{|z'_1|<4\lambda} \!\rmd z'_1\,
\frac{1}{|\alpha_2 -\omega(x_2)-|\nabla\omega(x_2)| z'_1+\ci\beta|} 
\nonumber \\ & \qquad\times 
\sup_{\substack{|z|,|z'|< 2\lambda \\ z_2,z'_2=0}}
\, \int_{t_1^2<(2\lambda)^2-z^2} \!\rmd t_1 
\int_{t_2^2<(2\lambda)^2-(z')^2}\rmd t_2\,
\frac{1}{|\alpha_3-f(t_1,t_2;z',z)+\ci\beta|}
\end{align}
where
\begin{align}
& f(t_1,t_2;z',z) = \omega(x_3+\gamma(O_2^T (z'+t_2 e_2),O_1^T (z+t_1 e_1)))
\nonumber \\ & \quad
= \omega(x_3+\gamma(\tilde{z}_2+t_2 \tilde{v}_2,\tilde{z}_1+t_1 \tilde{v}_1)) ,
\end{align}
with $\tilde{z}_2= O_2^T z'$ and $\tilde{z}_1=O_1^T z$.
Let us denote the final supremum by $J''$.  Applying Proposition 
\ref{th:genfboundn1}, we find the bound
\begin{align}\label{eq:Jpest}
& J'(\sigma)  
\le \frac{2^{4 d-6} N_d^2}{N_{d-2}^2} 
\frac{6^2 \sabs{\ln\sabs{\norm{\omega}'_1}}^2}{
    |\nabla\omega(x_1)|\, |\nabla\omega(x_2)| } 
\frac{\sabs{\ln \beta}^2}{\lambda^4} J'' .
\end{align}

We still need to estimate $J''$.  To do this we need to make a diversion
and first prove the following Lemma:
\begin{lemma}\label{th:nunlemma}
Let an integer $n\ge 2$ and 
$a,b,c\in \R$ be given and suppose $|c|\ge \vep'>0$.  Then there is
$\nu \in \R$, with $|\nu|,|1-\nu|\le 2$, for which
\begin{align}
 |\nu^n a + (1-\nu)^n b + c| \ge \frac{\vep'}{2}.
\end{align}
\end{lemma}
\begin{proof}
Let $f(\nu) = \nu^n a + (1-\nu)^n b + c$, and let 
us first assume that $a\ge |b|$.  Then, if $a\le |c|-\vep'/2$, we have
$|f(1)|\ge |c|-|a| \ge \vep'/2$, and choosing $\nu=1$ suffices.
Alternatively, 
if $a> |c|-\vep'/2$, we have $a>\vep'/2$ and thus
\begin{align}
 f(2) \ge 2^n a - |b| - |c|  
 >  a (2^n-2)-\frac{\vep'}{2}
 \ge  \frac{\vep'}{2} (2^n-3) \ge  \frac{\vep'}{2}.
\end{align}
Therefore, the estimate holds then for either $\nu=1$ or $\nu=2$.
If $a\le -|b|$, we have $-a\ge |b|$, and after swapping the signs of 
$a,b,c$, we can apply the above result to conclude that again
$|f(\nu)|\ge \vep'/2$ at either $\nu=1$ or $\nu=2$.  We have then proven the
result for $|a|\ge |b|$.  Finally, if $|a|<|b|$, we apply the above result
for $\nu'=1-\nu$, and conclude that in this case either $|f(0)|$ or
$|f(-1)|$ is greater than or equal to $\vep'/2$.
Thus the bound is attained at one of the points $\nu\in \set{-1,0,1,2}$,
which implies $|\nu|,|1-\nu|\le 2$.
\end{proof}

For both $j=1,2$, let $\barg_{j,n}=\barg_{n}(x_j,v_j)$ be defined 
as in item \ref{it:defbarg} of Lemma \ref{th:psiderivlemma}.  
Then we employ Lemma \ref{th:nunlemma} with $n=N$ and
\begin{align}
 a & = -\sigma_1^{N+1} \barg_{1,N} u_1\cdot \nabla \omega(x_3),
 \nonumber \\
 b & = -\sigma_2^{N+1}  \barg_{2,N} u_2\cdot \nabla \omega(x_3),
 \nonumber \\
 c & =  \frac{1}{N!} (v_3 \cdot \nabla)^{N} \omega(x_3) .
\end{align}
As $|c|> \frac{1}{2}\vep$, this yields a $\nu\in\R$ 
such that $|\nu|,|1-\nu|\le 2$, and $|\tilde{c}| \ge \frac{1}{4}\vep$ with
\begin{align}
& \tilde{c}=
\left( -\nu^{N}\sigma_1^{N+1} \barg_{1,N}u_1
 -(1-\nu)^N \sigma_2^{N+1} \barg_{2,N}  u_2\right)\cdot \nabla \omega(x_3)
 \nonumber \\ & \qquad 
+  \frac{1}{N!} (v_3 \cdot \nabla)^{N} \omega(x_3) .
\end{align}

Now $|\tilde{z}_1|= |z|<2\lambda$, $|\tilde{z}_2|= |z'|<2\lambda$, 
and also $\tilde{z}_j\cdot \tilde{v}_j=0$, with the $t_j$-integration 
going over values with $|t_j|^2<(2\lambda)^2-|\tilde{z}_j|^2$.  
We make the final change of variables $(t_1,t_2)\to (t,t')$, given by
\begin{align}
  t_1 = \sigma_1 (- t' + \nu t),\qand 
  t_2 = \sigma_2 (t' + (1-\nu) t)
\end{align}
where $\nu$  is the constant found above.
The Jacobian of the transformation is always $1$, and
it has the inverse
\begin{align}
  t = \sigma_1 t_1 +
  \sigma_2 t_2 \qand
  t' = \nu \sigma_2 t_2 - (1-\nu) \sigma_1 t_1,
\end{align}
and thus 
inside the new integration region
\begin{align}
|t|  \le |t_1|+|t_2| < 2^3 \lambda \qand
|t'| < (|\nu|+|1-\nu|) 4 \lambda \le  2^4 \lambda.
\end{align}
Therefore,
\begin{align}\label{eq:introttF}
& \int_{t_1^2<(2\lambda)^2-z^2} \!\rmd t_1 
\int_{t_2^2<(2\lambda)^2-(z')^2}\rmd t_2\,
\frac{1}{|\alpha_3-f(t_1,t_2;z',z)+\ci\beta|}
\nonumber \\ & \quad
\le 
\int\! \rmd t'
\int_{I(t')} \rmd t \,
\frac{2}{|\alpha_3- F(t;t')+\ci\beta|}
\nonumber \\ & \quad
\le 2^6 \lambda
\sup_{t'}
\int_{I(t')} \rmd t \,
\frac{1}{|\alpha_3- F(t;t')+\ci\beta|} 
\end{align}
where the integration region over $t$, that is $I(t')$, depends on $t'$, 
but it always is an interval of a length less than $2^4 \lambda$.
The final integral contains the function
\begin{align}
& F(t;t') =  f(t_1,t_2;z',z) 
= \omega(x_3+\gamma(\tilde{z}_2+t_2 \tilde{v}_2,\tilde{z}_1+t_1 \tilde{v}_1))
\nonumber \\ & \quad
=  \omega(\tilde{x}_3 + 
\sigma_1 (\psi_1(\tilde{y}_1 + \sigma_1 \nu t \tilde{v}_1)-
  \psi_1(\tilde{y}_1) ) 
\nonumber \\ & \qquad\quad
+ \sigma_2 (\psi_2(\tilde{y}_2 + \sigma_2 (1-\nu) t \tilde{v}_2)-
  \psi_2(\tilde{y}_2) ) )
\end{align}
where
\begin{align}
\tilde{y}_1 = \tilde{y}_1(t') =\tilde{z}_1-t'\sigma_1 \tilde{v}_1
\qand
\tilde{y}_2 =\tilde{z}_2+t'\sigma_2 \tilde{v}_2,
\end{align}
and 
\begin{align}
\tilde{x}_3 = \tilde{x}_3(t') = x_3 + 
\sigma_1 (\psi_1(\tilde{y}_1)-x_1) + 
\sigma_2 (\psi_2(\tilde{y}_2)-x_2) .
\end{align}

Let us then define, as in Proposition \ref{th:psiderivprop},
\begin{align}
 \gamma_j(\tau)  = \psi_j(\tilde{y}_j + \tau \tilde{v}_j) \qand
 \Gamma_j(\tau)  =  \gamma_j(\tau)-  \tau v_j- \psi_j(\tilde{y}_j).
\end{align}
As $\sigma_1 v_1 (\sigma_1\nu t) + 
\sigma_2 v_2 (\sigma_2(1-\nu) t) = (\nu v_1+ (1-\nu)v_2) t$, then
$F(t) = \omega(\Gamma(t))$ with
\begin{align}
&\Gamma(t) = \tilde{x}_3 +  t v_0 + 
\sigma_1\Gamma_1(\sigma_1\nu t) +
\sigma_2\Gamma_2(\sigma_2(1-\nu) t) .
\end{align}
Here $v_0 = \nu v_1+ (1-\nu)v_2$, and it thus satisfies
\begin{align}\label{eq:v0minusv3}
 |v_0-v_3| \le 2 ( |v_1-v_3| + |v_2-v_3| ) < 4 \delta.
\end{align}

By Lemma \ref{th:compdiff}, 
\begin{align}\label{eq:FNcder}
& \frac{1}{N!}\left|\frac{\rmd^N}{\rmd t^N}F(t)\right|
\ge \frac{1}{N!}
  \left|\Gamma^{(N)}(t)\cdot \nabla\omega(\Gamma(t))
 + (\Gamma^{(1)}(t)\cdot \nabla)^N \omega(\Gamma(t))\right|
\nonumber \\ & \qquad
 - \sum_{k=2}^{N-1} \sum_{m\in \N_+^k} \1\Bigl(\sum_{j=1}^k m_j = N\Bigr)
 \norm{\omega}'_k  \prod_{j=1}^k\! \left[ 
 \smash{\frac{1}{m_j!}}|\Gamma^{(m_j)}(t)| \right] .
\end{align}
Here $\Gamma^{(1)}(t) = v_0 + \nu \Gamma_1^{(1)}(\sigma_1\nu t) +
(1-\nu)\Gamma_2^{(1)}(\sigma_2(1-\nu) t)$, and, by 
(\ref{eq:v0minusv3}) and Proposition
\ref{th:psiderivprop}, it has the bound
\begin{align}\label{eq:G1v3diff}
  |\Gamma^{(1)}(t)-v_3| \le 4 \delta + 4 \vep \mu^{N} \le 1
\end{align}
which implies in particular that $|\Gamma^{(1)}(t)|\le 2$.
Note that we can apply the Proposition, since
$\vep$, $\mu$, and $N$ are clearly in the right range, and also
the expansion radius satisfies
$2\lambda \le \frac{1}{2}a_0^{-1} \vep_0 (r_0 \mu)^{n_0}
= a_0^{-1} \vep (r_0 \mu)^{N}$.
For all $n\ge 2$, we similarly get
\begin{align}
\Gamma^{(n)}(t) = \sigma_1^{n+1} \nu^n \Gamma_1^{(n)}(\sigma_1\nu t) +
\sigma_2^{n+1} (1-\nu)^n\Gamma_2^{(n)}(\sigma_2(1-\nu) t) 
\end{align}
satisfying, with $\tilde{C}=1+\norm{\omega}'_{n_0}\ge
1+\frac{\norm{\omega}'_N}{N!}$, the bounds
\begin{align}\label{eq:DGbignbounds}
\Bigl| \frac{1}{n!} \Gamma^{(n)}(t) \Bigr|
\le \begin{cases}
  2^{n+2} \vep \mu^{N-n} r_0^{N-1-n}, & \text{ for }2\le n<N\\
 2^{N+2} \tilde{C} r_0^{-1}, & \text{ for }n=N\\
  2^{N+3} \tilde{C} \mu^{-1}  r_0^{-2}, & \text{ for }n=N+1
 \end{cases} .
\end{align}

Consider then the sum over $k$ in (\ref{eq:FNcder}).  Since $k\ge 2$,
inside the sum always $m_j\le N-1$. Denoting, as before,
$\ell=\left|\defset{j}{m_j=1}\right|\le k-1$, we thus have
\begin{align}
& \prod_{j=1}^k\! \left[ 
 \smash{\frac{1}{m_j!}}|\Gamma^{(m_j)}(t)| \right] \le
2^{\ell + 3 (k-\ell) +\sum_{j} (m_j-1)} (\vep \mu^{N-1})^{k-\ell} \mu^{\sum_j (1-m_j)}
\nonumber \\ & \quad
 \le 2^{2 k + N} \vep \mu^{N-1+k-N} \le 2^{3 N-2} \vep \mu .
\end{align}
Therefore, the sum over $k$ is bounded by
\begin{align}
2^{3 N-2} \vep \mu \sum_{k=2}^{N-1} \binom{N-1}{k-1} \norm{\omega}'_{N-1}
\le 2^{4 N-3} \vep \mu \norm{\omega}'_{N-1} \le \frac{1}{2^4}\vep.
\end{align}

To estimate the first term in (\ref{eq:FNcder}), we use the estimates in item
\ref{it:bargbounds} of Lemma \ref{th:psiderivlemma}, with 
$b=\mu^{-1}\ge 1+ 2^{N} + \norm{\omega}'_{N+1} 2^{2 N+1}$.
Firstly,
\begin{align}\label{eq:FNcderp1}
&  \Bigl|\frac{1}{N!}\Gamma^{(N)}(t)\cdot \nabla\omega(\Gamma(t))
\nonumber \\ & \qquad
   - \left(-
 \nu^{N}\sigma_1^{N+1} \barg_{1,N}u_1 - (1-\nu)^N \sigma_2^{N+1} \barg_{2,N} u_2
\right)\cdot \nabla \omega(x_3) \Bigr|
\nonumber \\ & \quad
\le   \Bigl|\frac{1}{N!}\Gamma^{(N)}(t)\Bigr| \norm{\omega}'_2
|\Gamma(t)-x_3| 
\nonumber \\ & \qquad
+ |\nabla\omega(x_3)| 2^N 
\left(|g_{1,N}(\sigma_1\nu t)-\barg_{1,N}|  + 
|g_{2,N}(\sigma_2(1-\nu) t)-\barg_{2,N}|\right)
\nonumber \\ & \quad
\le 2^4 \lambda \norm{\omega}'_2 2^{N+2} \tilde{C} r_0^{-1}
  + \norm{\omega}'_1 2^{N+1} a_0 \mu^{1-N} \lambda r_0^{-N}
\nonumber \\ & \quad
\le 2^{N+3} \vep (r_0\mu)^N \tilde{C} r_0^{-1}
  + \norm{\omega}'_1 2^{N+1} \vep \mu
\le \vep \mu \left( \mu 2^{N+3} \tilde{C}
  + \norm{\omega}'_1 2^{N+1}\right)
\nonumber \\ & \quad
\le \vep \mu \left( 1 + \norm{\omega}'_1 2^{n_0+1}\right)
\le 2^{-n_0-2}\vep \le 2^{-4}\vep.
\end{align}
Secondly, by (\ref{eq:G1v3diff})
\begin{align}\label{eq:FNcder3}
& \left|(\Gamma^{(1)}(t)\cdot \nabla)^N \omega(\Gamma(t))
 - (v_3\cdot \nabla)^N \omega(x_3)\right|
\nonumber \\ & \quad
\le \sum_{k=1}^N \binom{N}{k} |\Gamma^{(1)}(t)-v_3|^k \norm{\omega}'_{N}
 + \norm{\omega}'_{N+1} 2^4 \lambda
\nonumber \\ & \quad
\le 4(\lambda + \vep \mu^N) \norm{\omega}'_{N} 2^N 
 + \norm{\omega}'_{N+1} 2^4 \lambda
\nonumber \\ & \quad
\le \vep \mu^N \norm{\omega}'_{N+1} ( 2^{N+3} +2^4 )
\le \vep \mu^2 \norm{\omega}'_{N+1} 2^{N+4}
\le 2^{-4} \vep.
\end{align}
Therefore,  
\begin{align}
& 
  \left| \frac{1}{N!}\Gamma^{(N)}(t)\cdot \nabla\omega(\Gamma(t))
 +  \frac{1}{N!}(\Gamma^{(1)}(t)\cdot \nabla)^N \omega(\Gamma(t))
 -\tilde{c}\, \right| \le 2^{-3} \vep
\end{align}
which can be combined with the previous estimate for the sum over $k$ in
(\ref{eq:FNcder}) to prove that for all allowed $t$
\begin{align}\label{eq:FNcderlast}
& \frac{1}{N!}\left|\frac{\rmd^N}{\rmd t^N}F(t)\right|
\ge |\tilde{c}| - \frac{3}{2^4} \vep \ge \frac{1}{2^4} \vep .
\end{align}

On the other hand, applying the Lemma \ref{th:compdiff}  
once more proves that
\begin{align}\label{eq:FNplus1}
& \frac{1}{(N+1)!}\left|F^{(N+1)}(t)\right|
\le 
\sum_{k=1}^{N+1} \sum_{m\in \N_+^k} \1\Bigl(\sum_{j=1}^k m_j = N+1\Bigr)
 \prod_{j=1}^{k-1} \frac{m_j}{\sum_{j'=j}^{k} m_{j'}}
\nonumber \\ & \qquad
\times \prod_{j=1}^k\!\Bigl[ \frac{1}{m_j!}|\Gamma^{(m_j)}(t)| \Bigr] \!
\norm{\omega}'_k .
\nonumber \\ & \quad
\le \frac{|\Gamma^{(N+1)}(t)| }{(N+1)!}
 + \tilde{C} r_0^{-1} \sum_{k=2}^{N+1} \sum_{m\in \N_+^k} 
\1\Bigl(\sum_{j=1}^k m_j = N+1\Bigr)
2^{\sum_j (m_j+2)} \norm{\omega}'_{N+1} 
\nonumber \\ & \quad
\le   2^{N+3} \tilde{C} \mu^{-1}  r_0^{-2}
 + \tilde{C} r_0^{-1} \sum_{k=2}^{N+1} \binom{N}{k-1} 2^{N+1+2 k} 
 \norm{\omega}'_{N+1} 
\nonumber \\ & \quad
\le  2^{N+3} \tilde{C} \mu^{-1}  r_0^{-2} \left( 1
 + \mu 2^{3 N} \norm{\omega}'_{N+1}  \right)
\le  2^{N+4} \tilde{C} \mu^{-1}  r_0^{-2} 
\end{align}
where we have used the fact that $m_j=N$ can appear only once in the product
over $j$.  Since $r_0\ge \frac{1}{2} \beta^\gamma$, we have also
\begin{align}
& \vep'= 
\frac{\vep}{2^4} \frac{1}{2^{N+1}(N+1)^N} \frac{r_0^2\mu}{2^{N+4} \tilde{C}}
\ge \vep_0 r_0^{n_0-N+2}
 \frac{\mu^{n_0-N+1}}{\tilde{C}} \frac{1}{2^{2 n_0 +8}(n_0+1)^{n_0}} 
\nonumber \\ & \quad
\ge \beta^{\gamma n_0} \vep_0 
 \frac{\mu^{n_0-1}}{\tilde{C}} \frac{1}{2^{3 n_0 +8}(n_0+1)^{n_0}} .
\end{align}
If this is raised to the power $N+1$, the result is bounded from below 
by an ($n_0$-dependent) constant times $\beta^{\gamma n_0 (n_0+1)}$.
Therefore, as long as $\gamma^{-1} > n_0 (n_0+1)$, there is $\beta_0>0$,
such that we can apply 
the conclusion of Proposition \ref{th:genfbound} for all 
$0<\beta\le \beta_0$.  For such values of $\beta$ and all
allowed $t'$, we have
\begin{align}\label{eq:Jppend}
& \int_{I(t')}\! \rmd t \,
\frac{1}{|\alpha_3- F(t;t')+\ci\beta|} 
\nonumber \\ & \quad
 \le 2^{N+1}(N+1)^N \left( \frac{2^4 \lambda}{\vep 2^{-4}}
   \beta^{\frac{1}{N+1}-1} +
 2^{N+4} \tilde{C} \mu^{-1}  r_0^{-2} 
 2^{\frac{4}{N}} \vep^{-\frac{1}{N}} \beta^{\frac{1}{N}-1} \right) 
\nonumber \\ & \quad
\le 2^{n_0+1}(n_0+1)^{n_0} \Bigl(2^8 \frac{\vep_0}{4 a_0} (r_0 \mu)^{n_0}
 \frac{1}{2 \vep_0} (r_0 \mu)^{N-n_0} \beta^{\frac{1}{N+1}-1} 
\nonumber \\ & \qquad
+ 2^{n_0+6} \tilde{C} \mu^{-1} r_0^{-2} 
\left(2 \vep_0 (r_0 \mu)^{n_0-N}\right)^{-\frac{1}{N}}
 \beta^{\frac{1}{N}-1} \Bigr) 
\nonumber \\ & \quad
\le 2^{3 n_0+8}(n_0+1)^{n_0}  \tilde{C} \mu^{-\frac{n_0}{N}} \vep_0^{-\frac{1}{N}}
 \Bigl( \beta^{\frac{1}{N+1}-1} + \beta^{-\gamma (1+\frac{n_0}{N})
   +\frac{1}{N}-1} \Bigr) .
\end{align}
Since we have not aimed at optimal estimates here, we do not try to
optimize the extra decay arising from the crossing.  Instead, let us prove
that the choice given in (\ref{eq:defgammafin}) is sufficient.
Then we can also choose explicitly
\begin{align}\label{eq:defbeta0}
  \beta_0 = \Bigl( \vep_0 
 \frac{\mu^{n_0-1}}{\tilde{C}} \frac{1}{2^{3 n_0 +8}(n_0+1)^{n_0}}
 \Bigr)^{\frac{3}{2} (n_0+1)} 
\end{align}
since, for all $0<\beta\le \beta_0$, then 
$\beta \le \beta^{1/3}  \beta_0^{2/3}\le (\vep')^{N+1}$.

With these choices, the power of $\beta$ in the second term in
(\ref{eq:Jppend}) is 
\begin{align}
& -\gamma (1+\frac{n_0}{N}) +\frac{1}{N}-1 = 
\frac{1}{3 n_0 (n_0+1) N} (-N - n_0 +3 n_0+ 3 n_0^2)-1
\nonumber \\ & \quad
\ge \frac{n_0}{(n_0+1) N}-1
\ge \frac{1}{n_0+1}-1 .
\end{align}
Therefore, by (\ref{eq:introttF}), we have proved
\begin{align}
& J'' \le \lambda 2^{3 n_0+15}(n_0+1)^{n_0}  \tilde{C} \mu^{-\frac{n_0}{2}}
\vep_0^{-\frac{1}{2}}  \beta^{\frac{1}{n_0+1}-1} .
\end{align}
Combining this with (\ref{eq:Jpest})
proves the validity of (\ref{eq:JCp2bound}) for 
\begin{align}
  C'_2 = 2^{3 n_0+9+4 d}(n_0+1)^{n_0}  \tilde{C} \mu^{-\frac{n_0}{2}}
  \vep_0^{-\frac{1}{2}} \frac{N_d^2}{N_{d-2}^2} 
  6^2 \sabs{\ln\sabs{\norm{\omega}'_1}}^2 \norm{\omega}'_1
\end{align}
when $\gamma$ is chosen as in (\ref{eq:defgammafin}) and $\beta$
is sufficiently small.  For notational simplicity, we have added the missing
gradient factor $|\nabla\omega(x_3)|$ to the denominator: this makes the
estimate invariant under permutations of the indices, and thus allows to use
it directly for the original integral.
This completes the proof of Theorem \ref{th:thmain}.

\appendix

\section{Differentials of composite functions}
\label{sec:composite}

\begin{lemma}\label{th:compdiff} 
Let $d,n\in \N_+$, an open interval $I$,
and $\Gamma\in C^{(n)}(I,\R^d)$, and $f\in C^{(n)}(\R^d,\R)$ be given. 
Then for all $t\in I$, 
\begin{align}\label{eq:combder}
& \frac{1}{n!}\frac{\rmd^n}{\rmd t^n}f (\Gamma(t))
= \sum_{k=1}^n \sum_{m\in \N_+^k} \1\Bigl(\sum_{j=1}^k m_j = n\Bigr)
 \prod_{j=1}^{k-1} \frac{m_j}{\sum_{j'=j}^{k} m_{j'}}
\nonumber \\ & \qquad
\times \prod_{j=1}^k\! \left.\left[ 
 \smash{\frac{1}{m_j!}}\Gamma^{(m_j)}(t) \cdot \nabla \right] \!
f \right|_{\Gamma(t)} .
\end{align}
\end{lemma} 
\begin{proof}
The result holds for $n=1$.  For the induction step, let us assume it holds 
for values up to $n$.  Then 
\begin{align}
& \frac{\rmd}{\rmd t}
\frac{1}{n!}\Bigl[\frac{\rmd^{n}}{\rmd t^n}f (\Gamma(t))\Bigr]
= \sum_{k=1}^n \sum_{m\in \N_+^k} \1\Bigl(\sum_{j=1}^k m_j = n\Bigr)
 \prod_{j=1}^{k-1} \frac{m_j}{\sum_{j'=j}^{k} m_{j'}}
\nonumber \\ & \qquad
\times \Bigl\{
\sum_{\ell=1}^k
 \prod_{j=1,j\ne \ell}^k\! \left.\left[ 
 \smash{\frac{1}{m_j!}}\Gamma^{(m_j)}(t) \cdot \nabla \right] \!
\left[ 
 \smash{\frac{1}{m_\ell!}}\Gamma^{(m_\ell+1)}(t) \cdot \nabla \right]
f \right|_{\Gamma(t)} 
\nonumber \\ & \qquad\quad
+  \prod_{j=1}^k\! \left.\left[ 
 \smash{\frac{1}{m_j!}}\Gamma^{(m_j)}(t) \cdot \nabla \right] \!
\left[ \Gamma^{(1)}(t) \cdot \nabla \right]
f \right|_{\Gamma(t)} 
\Bigr\}
\end{align}
In the first term,
we take out the sum over $\ell$, and then change variables from $m$ to
$M$ so that $M_{\ell} = m_\ell +1$ and otherwise $M_j = m_j$.
This yields a term
\begin{align}\label{eq:cd1st}
&  \sum_{k=1}^n \sum_{\ell=1}^k
\sum_{M\in \N_+^k} \1\Bigl(\sum_{j=1}^k M_j = n+1\Bigr) \1(M_\ell\ge 2)
 \prod_{j=\ell+1}^{k-1} \frac{M_j}{\sum_{j'=j}^{k} M_{j'}}
\nonumber \\ & \qquad
 M_{\ell} \frac{M_\ell-1}{\sum_{j'=\ell}^{k} M_{j'}-1}
\prod_{j=1}^{\ell-1} \frac{M_j}{\sum_{j'=j}^{k} M_{j'}-1}
 \prod_{j=1}^k\! \left.\left[ 
 \smash{\frac{1}{M_j!}}\Gamma^{(M_j)}(t) \cdot \nabla \right] \!
f \right|_{\Gamma(t)} .
\end{align}
For the second term, we add one more sum over $m=1$, and then shift the $k$
sum accordingly.  This yields
\begin{align}\label{eq:cd2nd}
&  \sum_{k=2}^{n+1}
\sum_{m\in \N_+^k} \1\Bigl(\sum_{j=1}^k m_j = n+1\Bigr) \1(m_k=1)
 \prod_{j=1}^{k-1} \frac{m_j}{\sum_{j'=j}^{k} m_{j'}-1}
\nonumber \\ & \qquad
 \prod_{j=1}^k\! \left.\left[ 
 \smash{\frac{1}{m_j!}}\Gamma^{(m_j)}(t) \cdot \nabla \right] \!
f \right|_{\Gamma(t)} .
\end{align}
It is then an explicit computation to check that the $k=1$ term in 
(\ref{eq:cd1st}) is equal to the $k=1$ term in (\ref{eq:combder}) times $n+1$
(after setting $n\to n+1$), 
and that the same holds for $k=n+1$ term in (\ref{eq:cd2nd}).

For $2\le k \le n$ we need to sum the corresponding terms in
(\ref{eq:cd1st}) and (\ref{eq:cd2nd}).  Their sum can be written as
\begin{align}\label{eq:cd3rd}
&  \sum_{m\in \N_+^k} \1\Bigl(\sum_{j=1}^k m_j = n+1\Bigr)
 \prod_{j=1}^{k-1} \frac{m_j}{\sum_{j'=j}^{k} m_{j'}}
 \prod_{j=1}^k\! \left.\left[ 
 \smash{\frac{1}{m_j!}}\Gamma^{(m_j)}(t) \cdot \nabla \right] \!
f \right|_{\Gamma(t)} .
\nonumber \\ & \quad
\times \Bigl\{ \sum_{\ell=1}^k \1(m_\ell\ge 2)
 (m_\ell-1) \prod_{j=1}^{\ell} 
\frac{\sum_{j'=j}^{k} m_{j'}}{\sum_{j'=j}^{k} m_{j'}-1}
\nonumber \\ & \qquad
 +  \1(m_k=1)  \prod_{j=1}^{k-1} 
\frac{\sum_{j'=j}^{k} m_{j'}}{\sum_{j'=j}^{k} m_{j'}-1}
\Bigr\} .
\end{align}
For the computation of the term in the curly brackets, let us separate
$\ell=k$ term.  When $\ell<k$, all the terms in the denominator are
non-zero, as for $j<k$, we have $\sum_{j'=j}^{k} m_{j'}-1>0$ due to
$m_\ell\ge 1$.
Therefore, we can apply the property
\begin{align}
 (m_\ell-1) \1(m_\ell\ge 2) =  m_\ell-1 = 
\sum_{j'=\ell}^{k} m_{j'} -1
- \sum_{j'=\ell+1}^{k} m_{j'} ,
\end{align}
which shows that the sum over $\ell< k$ is equal to
\begin{align}%
& \sum_{\ell=1}^{k-1}
\frac{\prod_{j=1}^{\ell} \sum_{j'=j}^{k} m_{j'}}{
  \prod_{j=1}^{\ell-1} (\sum_{j'=j}^{k} m_{j'}-1)}
- \sum_{\ell=1}^{k-1}
\frac{\prod_{j=1}^{\ell+1} \sum_{j'=j}^{k} m_{j'}}{
  \prod_{j=1}^{\ell} (\sum_{j'=j}^{k} m_{j'}-1)}
\nonumber \\ & \quad
 =   \sum_{j'=1}^{k} m_{j'}
 - m_{k} \prod_{j=1}^{k-1} \frac{\sum_{j'=j}^{k} m_{j'}}{
 \sum_{j'=j}^{k} m_{j'}-1} .
\end{align}
The second term here 
is canceled by the remaining terms in the curly brackets.
(If $m_k>1$, the $\ell=k$ term in the sum cancels it
and the last term in (\ref{eq:cd3rd}) is zero; if $m_k=1$, the  
opposite happens.)  Therefore, the term in the curly brackets is equal to
$\sum_{j=1}^{k} m_{j} = n+1$.  This proves (\ref{eq:combder}).
\end{proof}

\section{Properties of $\sabs{x}$}
\label{sec:sabsxprop}

\begin{proposition}
Let $\sabs{x}=\sqrt{1+x^2}$.  Then for all $x,y\in \R$,
\begin{enumerate}
\item\label{it:sabs1} $|x|< \sabs{x}$.
\item\label{it:sabs4} If $|x|\le |y|$, then $\sabs{x}\le \sabs{y}$ and
$\sabs{\ln \sabs{x}}\le \sabs{\ln \sabs{y}}$.
\item\label{it:sabs2} $\sabs{x+y}<\sabs{x}+\sabs{y}\le 2 \sabs{x}\sabs{y}$.
\item\label{it:sabs3} $\sabs{x y}\le \sabs{x}\sabs{y}$, and, if $|x|\ge 1$,
  $\sabs{x y}\le |x|\sabs{y}$. 
\end{enumerate}
\end{proposition}
\begin{proof}
Items \ref{it:sabs1} and \ref{it:sabs4} are obvious.  
The first inequality of item
\ref{it:sabs2} is proven by
\begin{align}
& \sabs{x}+\sabs{y}-\sabs{x+y} = 
\frac{(\sabs{x}+\sabs{y})^2-\sabs{x+y}^2}{\sabs{x}+\sabs{y}+\sabs{x+y}}
 \nonumber \\ & \quad
 = \frac{2 \sabs{x} \sabs{y} + 1+ x^2 + 1+y^2 - 
   (1+x^2+y^2+2 x y)}{\sabs{x}+\sabs{y}+\sabs{x+y}}
 \nonumber \\ & \quad
 = \frac{1 + 2 \sabs{x} \sabs{y} (1-\frac{x}{\sabs{x}}\frac{y}{\sabs{y}})}{
   \sabs{x}+\sabs{y}+\sabs{x+y}} >0.
\end{align}
The proofs of the remaining inequalities in  \ref{it:sabs2} and \ref{it:sabs3}
are very similar, and we will skip them.
\end{proof}

\section{Morse functions}
\label{sec:Morse}

We prove here the following result which shows that Morse functions are
covered by the main results given in text.
\begin{proposition}\label{th:morseisOK}
Let $d\ge 3$, and assume $\omega$ is a real-analytic and 
$\Z^d$-periodic Morse function on $\R^d$.  Then
$\omega$ satisfies Assumption \ref{th:DRass}, and we can take
$p_0=0$ for $d>4$ and $p_0=1$ for $d=3$.
\end{proposition}
\begin{proof}
Define $f_\omega$ by (\ref{eq:deffom}).  Let $X=[-1/2,1/2]^d$, and let $x_j$,
$j=1,\ldots,n$, enumerate the critical points of $\omega$ in $X$ (as
$\omega$ is a Morse function, there can be no accumulation of its
critical points, and thus $n<\infty$).  Let also $M_j=D^2 \omega(x_j)$
be the Hessian of $\omega$ at $x_j$, let 
$\lambda^{(i)}_{j}$ denote its eigenvalues, and define 
$a_j = \min_i |\lambda^{(i)}_{j}|$ and $b_j = \max_i |\lambda^{(i)}_{j}|$.
By assumption, $M_j$ is invertible, and thus we have 
$0<a_j \le b_j < \infty$.

By Taylor's formula, now for any $x\in\R^d$ and $j$,
\begin{align}
\nabla \omega(x) = \nabla \omega(x)-\nabla \omega(x_j)
= M_j (x-x_j) + R_j(x)
\end{align}
where $|R_j(x)|\le \frac{1}{2} \norm{\omega}'_3 |x-x_j|^2$.
Here, by using an orthogonal transformation which diagonalizes the 
Hermitian matrix $M_j$, we find
\begin{align}
a_j |x-x_j|\le |M_j (x-x_j)|\le b_j|x-x_j|.
\end{align}
Let $r_j = a_j/\norm{\omega}'_3$ which is non-zero, as 
$\norm{\omega}'_3$ is finite.  Then we can conclude, by using
the triangle inequality, that whenever $|x-x_j|\le r_j$,
\begin{align}
\frac{a_j}{2}|x-x_j|\le |\nabla \omega(x)| \le \frac{3 b_j}{2}|x-x_j|.
\end{align}

Let $U_j = \defset{x}{|x-x_j|<r_j}$, $j=1,\ldots,n$, and denote
$K=X\setminus (\cup_j U_j)$.  Then $K$ is compact, and contains
no critical points of $\omega$.  Therefore, by continuity of 
$\nabla \omega$, we have $\vep=\min_{x\in K}|\nabla \omega(x)|>0$.
By splitting the integration region into parts by removing the
balls $U_j$, we find that for all $0<s\le \vep$,
\begin{align}\label{eq:fomest}
& f_\omega(s) = 
\int_{X}\! \rmd x\, \frac{1}{|\nabla \omega(x)|^3} 
\1(|\nabla \omega(x)|\ge s)
 \nonumber \\ & \quad
\le \int_{K}\! \rmd x\, \frac{1}{\vep^3} + \sum_{j=1}^n  
\int_{U_j}\! \rmd x\, \frac{1}{|\nabla \omega(x)|^3} 
\1(|\nabla \omega(x)|\ge s)
 \nonumber \\ & \quad
\le  \frac{1}{\vep^3} + \sum_{j=1}^n  2^{-3} a_j^3 |S^{d-1}|
\int_{2 s/(3 b_j)}^{r_j}\! \rmd r\, r^{d-1-3} .
\end{align}
If $d>3$, then the final integral over $r$ is less than
$\int_{0}^{r_j}\! \rmd r\, r^{d-1-3} = \frac{1}{d-3} r_j^{d-3}$.
Therefore, we can conclude that then $\inf_{s>0} f_\omega (s) < \infty$,
as claimed in the Proposition.  Otherwise, $d=3$, and
\begin{align}
\int_{2 s/(3 b_j)}^{r_j}\! \rmd r\, r^{d-1-3}
= \ln \left(\frac{3 b_j r_j}{2 s}\right)
= \ln \left(\frac{3 b_j r_j}{2}\right) + \ln s^{-1}.
\end{align}
Then (\ref{eq:fomest}) implies that $f_\omega(s)\le c_0 \sabs{\ln s}$ for
some finite constant $c_0$.  This proves that also then Assumption 
\ref{th:DRass} is valid, this time with $p_0=1$.
\end{proof}


\end{document}